# Negative differential conductance in molecular junctions: an overview of experiment and theory

Bingqian Xu

*Single Molecule Study Laboratory, College of Engineering and Nanoscale Science and Engineering Center, University of Georgia, Athens, USA*

Yonatan Dubi

*Department of Chemistry and the Ilse Katz Center for nanoscale Science and Technology, Ben-Gurion University of the Negev, Beer Sheva, Israel*

**Abstract**

One of the ultimate goals of molecular electronics is to create technologies that will complement – and eventually supersede - Si-based microelectronics technologies. To reach this goal, electronic properties that mimic at least some of the electrical behaviors of today's semiconductor components must be recognized and characterized. An outstanding example for one such behavior is negative differential conductance (NDC), in which an increase in the voltage *across* the device terminals results in a decrease in the electric current passing *through* the device. This overview focuses on the NDC phenomena observed in metal-single molecule-metal molecular junctions, and is roughly divided into two parts. In the first part, the central experiments which demonstrate NDC in single-molecule junctions are critically reviewed, with emphasis on the main observations and their possible physical origins. The second part is devoted to the theory of NDC in single-molecule junctions, where simple models are employed to shed light on the different possible mechanisms leading to NDC.

## 1. Introduction

One of the ultimate goals of molecular electronics is to create technologies that are not only complementary to currently Si-based microelectronics technologies, but also that will eventually supersede them [1]. Exploiting the notion that molecules can be effectively wired to bulk electrodes, researchers must start by recognizing and promoting the electronic properties that mimic at least some of the electrical behaviors of today's semiconductor components. To that end, the functional current-voltage (I-V) behavior entailed in the negative differential conductance (NDC) effect of molecular junction systems has been extensively studied not only for its counterintuitive nature, but also due to its wide array of potential future applications [2-9].

Used to depict essentially non-monotonic current-voltage, NDC describes an uncommon property of certain electrical components, in which an increase in the voltage *across* the device terminals results in a decrease in the electric current passing *through* the device. In an ordinary resistor, in contrast, increases in bias always cause proportional increases in current. At the macroscopic level, the NDC effect is the principal feature of the I-V characteristics of resonant tunneling diodes, the bulk semiconductor devices pioneered by Esaki and his coworkers [10, 11]. In their work, the resonant tunneling of electrons was observed in a diode fabricated with double-barrier structures comprising a thin layer of small-gap material such as GaAs sandwiched between two GaAlAs layers with a wide gap between them (figure 1a) [10]. Resonant tunneling occurs when the carrier goes through an energy eigenstate of the well. However, the voltage bias not only shifts the chemical potentials, but also distorts the bands, inducing a shift in the position of the well resonant levels. The current maxima occur when the applied voltages to the barrier layers are such that the Fermi energy at the electrodes aligns with one of the states in the potential well (figure 1b), resulting in resonant tunneling which is manifested experimentally as peaks or humps in the tunneling current at voltages near the quasi-stationary states of the potential well. The NDC mechanism observed in Esaki's work established the theoretical basis and has been widely adopted in today's practical applications of bulk electronic devices, such as negative resistance oscillators, amplifiers and switching circuits [12-14].

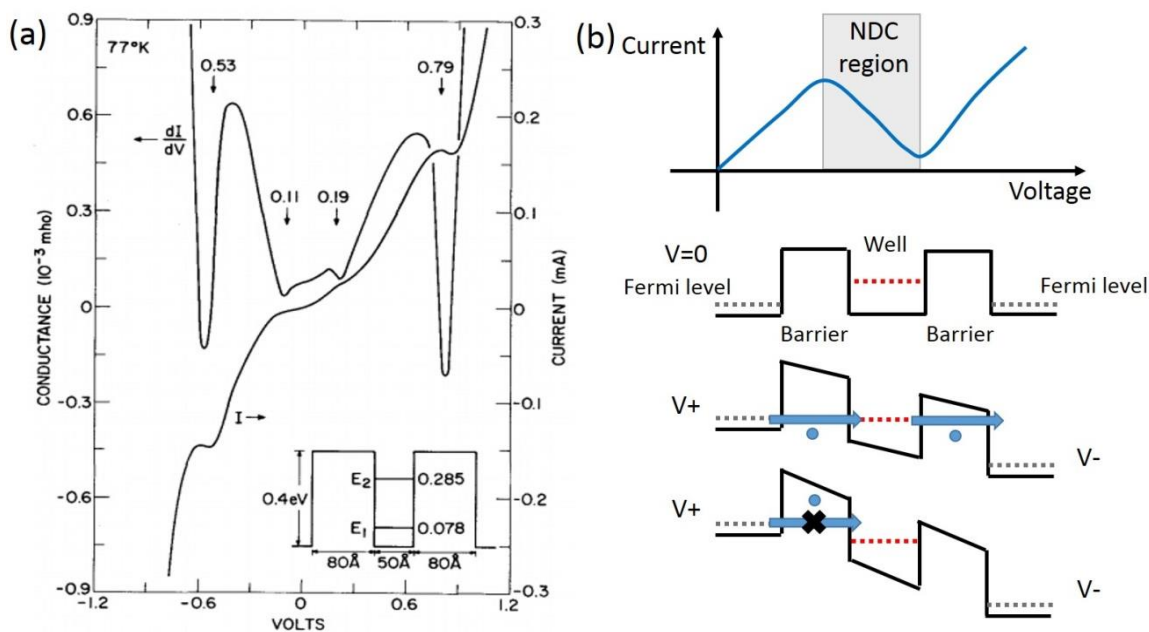

**Figure 1:** (a) Current and differential conductance characteristics of the resonant tunneling diode pioneered by Esaki and coworkers (used with permission from [11]). (a)The inset shows the energy diagram of the double-barrier structure. (b) A schematic illustration of NDC phenomena for resonant tunneling diode and the underlying mechanism responsible for it.

In the past decade and a half, observations of NDC at the single-molecule level have revived interest in the NDC phenomenon and stimulated global research efforts driven by the desire to apply molecular NDC in molecule-based nanotechnology, such as molecular junction systems where it has recently been observed [7, 15-18]. To mimic the mechanism of Esaki's resonant diode, molecular junctions comprising semiconductor materials as their electrodes – i.e., metal-molecule-semiconductor system – were developed to generate NDC phenomena [16, 19-24]. This approach exploits the presence of a band-edge of heavily doped semiconductor material that strongly modifies the electron injection rate, leading to NDC when the molecular level shifts into the gap at the band-edge [24].

Far from being a comprehensive review of NDC, this overview is limited in its focus to NDC in metal-molecule-metal junctions. Insofar as they are the ultimate limit of electronic nanotechnology [25], with potential applications that extend far beyond electronic transport [26, 27], such molecular junctions (MJs) have been rigorously studied over the past 15 years. As we will describe below in detail, NDC has been observed in MJs in many experiments and discussed in numerous theoretical studies. Our goal here, however, is not to provide a comprehensive survey of past studies, but rather, to focus on what we feel are the essential experiments and theories in this field, in the process providing the reader with a simple, physical picture of this fascinating phenomenon.

The structure of this paper is as follows. Section 2 is dedicated to a (partial) survey of the central experiments in the field divided according to experimental method. For each experiment, we focus on the central observations (and their possible physical origins) and address the study limitations. Section 3 is devoted to the theory of NDC in molecular junctions. Divided according to the various possible mechanisms for NDC, the section includes simple (almost toy) models that shed light on the different mechanisms. Each of the theoretical models presented is also associated with a relevant experiment discussed in Section 2. Finally, we conclude in Section 4 with some prospects for future work in the field.

## 2. NDC in metal-molecule-metal junctions: Experiments

Experimental attempts to observe NDC using molecular junctions comprising metallic leads have attracted great interest, and over the last decade this effect has been observed in a large variety of molecules, including organic and metallo-organic molecules and even DNA junctions [8, 17, 28-31]. NDC is characterized by two important factors that can vary substantially across experiments. The first is the NDC voltage, namely, the voltage at which the current reaches a maximum. The second is the so-called peak-to-valley ratio (PVR), which is the ratio between the maximal (peak) and the minimal (valley) currents. Practically speaking, the NDC voltage should be as low as possible to minimize device power consumption while the PVR should be as large as possible.

**2.1. Measuring NDC behavior using the nanopore technique**

The experimental determination of molecular junction NDC was pioneered by Reed and coworkers [2, 3, 32, 33], whose work at the turn of the century stimulated, in the fourteen years since, extensive experimental (and theoretical) exploration of this intriguing current-voltage behavior [2, 3, 32]. In the experiments of Reed et al., a series of single oligo(phenylene ethynylnene)s (OPE) molecules with different substitutions were studied (figure 2). OPE molecules are attractive targets for molecular electronics because 1) the relatively small HOMO-LUMO gap (~ 3eV) of the molecule confers on them efficient electron transport, and 2) the synthetic flexibility to alter their chemical moieties makes them good candidates for the study of substituent effects [34]. In the system of Reed et al., one terminal of each OPE molecule was thiolated to make contact with one of the metallic electrodes. The molecules were measured using a "nanopore" technique (figure 3a) that, it should be pointed out, involves the insertion of a large number of molecules into a junction structure, only one side of the contact interfaces of which is well defined by the covalent bond between the thiol group and the metal lead. In a study that used the "nanopore" technique, OPE molecules substituted asymmetrically by π-active groups with nitroamine redox centers (molecule 1 in figure 2) were reported to have a negative differential resistance (NDR) feature at an external voltage of around 2.2 V when measured at 60 K in a high vacuum (figure 3b) [2]. Measurements under this condition yielded a very sharp peak-to-valley transition with a PVR of 1030:1, the highest PVR of NDC reported within metal-molecule-metal junction systems. However, the peak current value and peak voltage show opposite dependencies on the increase in temperature, and the PVR nearly vanishes when the temperature reaches room temperature [2].

Reed and colleagues performed further experiments at room temperature with several other molecules, and these also exhibited NDC features with relatively low PVR values of less than 2 [3, 32]. The observed degradation in NDC behavior with the increase in temperature is believed to be due to the increase in inelastic scattering at higher temperatures. NDC was also reported for the same conjugated backbone substituted only by the nitro group (molecule 2 in figure 2) at room temperature with a PVR of around 2:1 [3, 5]. In contrast, neither the backbone substituted only with the amino group on its central ring nor the unsubstituted backbone displayed NDC behavior [3]. These measurements highlighted the critical role of the redox center in generating the NDC signatures of molecular junction systems.

To explain the peak profiles of the substituted OPE molecules (molecule 1 and 2), two mechanisms have been proposed: 1) the NDC is related to a sharp conformational change associated with the twisting of the central ring due to the interaction between the electric field and the permanent dipole moment of the molecules, and 2) the electronic delocalization of the LUMO level of the molecules in their singly charged state is responsible for the current peak in the I/V characteristics [33, 35].

By utilizing strong σ-bonds (such as $CH_2$ linkages in an oligomer structure), one can effectively introduce tunnel barriers into the junction. The transport barriers are $CH_2$ because alkyl units pose a larger electronic transport barrier in conjugated moieties [32]. Using the nanopore technique, NDC was also observed to appear at ~1.7 V for such a molecular structure (molecule 4 in figure 2) at room temperature with a PVR of ~1.5 [32]. However, given the great number of molecules involved in the junction structure constructed using the nanopore technique, it is difficult to determine how many of the molecules are actually being measured or how many are responsible for the resulting NDC behavior. Thus, the NDC measured in the system of Reed et al. may be elicited by the collective behavior of multiple molecular junctions. In addition, the asymmetrical definitions of the two junction contact interfaces at the molecule-electrode interfaces add further uncertainty to the system. Noticeably, the active NDC voltage for these substituted OPE molecules is relatively high (> 1.5 V), which suggests that they will also have correspondingly high power consumption.

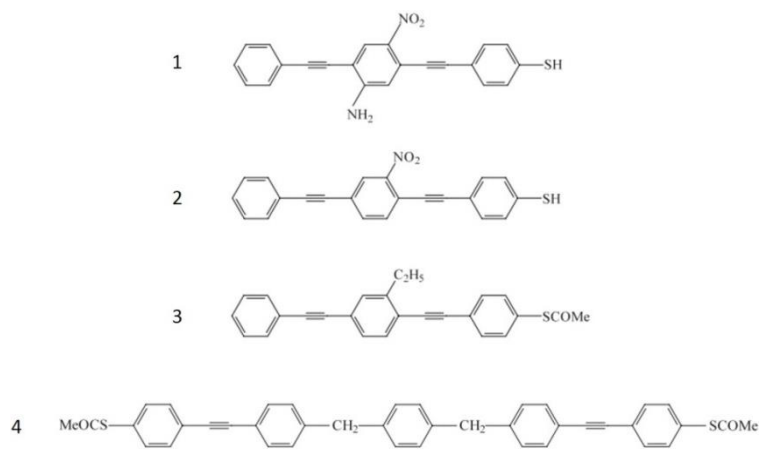

Figure 2: OPE molecules (1, 2, and 3) and other molecule (4) in the studies by Reed et al.

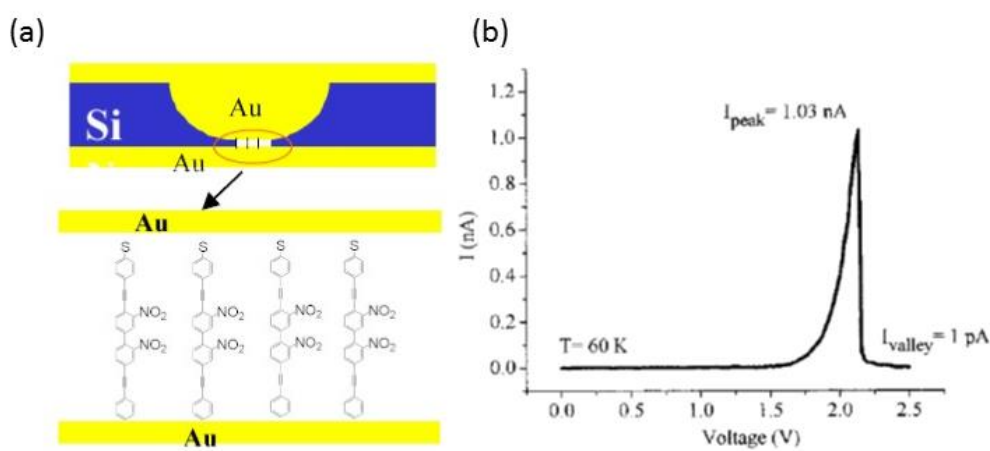

Figure 3: Au-SAM-Au junction formed by the nanopore technique (a) in a nanopore area and its I-V characteristics (b). Adapted with permission from [2].

## 2.2 Measuring NDC behavior using the STM technique

NDC behavior can also be revealed using a scanning tunneling microscopy (STM) setup [4, 7, 15, 36-38]. It is necessary to note that the STM set up mainly involves two regimes: 1) STM tip in contact with target molecules [4], and 2) tip without touching target molecules [15, 39, 40]. In figure 4a, molecular junction structure is constructed with a STM tip touching a self-assembled monolayer of target molecules. In contrast to measurements made using the nanopore technique, this STM setup exploit the apparatus' conducting tip as one of the electrodes and a metallic substrate as the other, a scenario in which, compared to the nanopore technique, fewer molecules in the junction structure need to be tested than for the nanopore technique due to the relatively smaller contacting area of a STM tip. Using this setup, self-assembled monolayers (SAM) made of molecules 2 and 3 (figure 2) were reported to exhibit NDC behavior at around 3 V (for both molecules) when adsorbed on the gold surface and in contact with a platinum (Pt) STM tip (figure 4b) [4]. Using the same setup, SAM of 4-p-terphenylthiol molecules adsorbed on a gold (111) surface and in contact with a Pt STM tip also displayed tip-induced NDC behavior at a voltage magnitude of ~3.5 V under both bias polarities [7].

However, the mechanism responsible for the NDC signal detected using the nanopore method cannot readily be applied to the STM experiments because of the difference in junction structure. For some cases, the narrow density of states (DOS) at the STM tip were proposed to be the cause for the observed NDC, and as such, the abrupt drop in voltage at the tip–molecule interface may be a critical determinant of I/V curve shape [4, 7, 41]. Interestingly, measurements of molecule 2 showed – under both the nanopore and STM measurement protocols at room temperature – that it has NDC behavior, but the amplitude of the active NDC voltage varied from 1 V measured using the nanopore technique to 2.3 V when measured using STM. In addition, the PVR value for NDC in the STM measurement scenario was much higher than that in the nanopore measurement environment [3, 4]. Thus, despite the significant role played by the nitro redox center, which is supposedly the principal reason that molecule 2 exhibits NDC in the nanopore measurements, the presence of the STM tip in STM junctions has a marked effect on the energy potential profile of the molecules being measured. Note, for example, that the active voltages associated with the NDC behaviors detected by STM measurements made at room temperature are still relatively high. More importantly, in the STM set up with tip in contact with molecules, the number of molecules being measured is typically unknown.

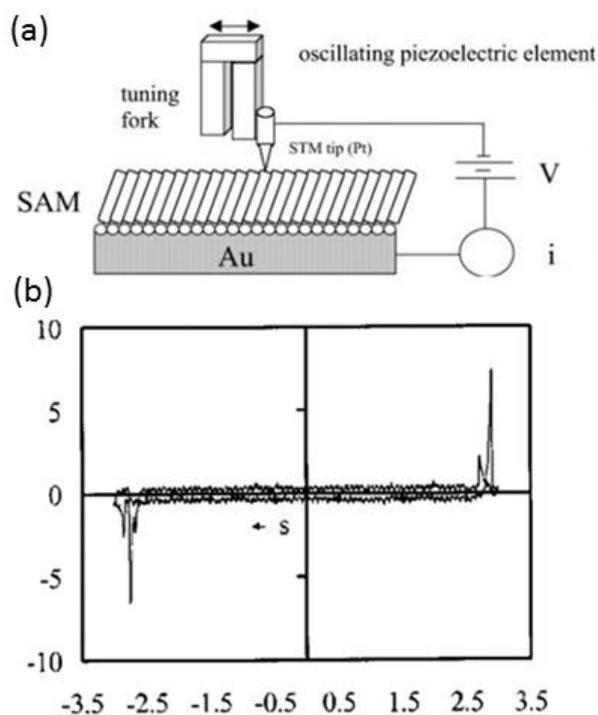

**Figure 4:** (a) Schematic of the STM measurement setup. (b) I-V curves of molecule 3 in figure 3 measured with STM technique. Adapted with permission from [4].

Using the STM setup when the tip is maintained at a tunneling distance without touching the molecule, some experiments performed at low temperatures and under high vacuum conditions have shown promising NDC features [15]. For example, pyrrolidine ($C_4H_8NH$) molecules in a junction comprising an STM tip and Cu(001) surface measured at low temperature (9 K) and under high vacuum conditions exhibited NDC behavior [15]. Interestingly, unlike the high active NDC voltage found in other experiments carried out in high vacuums, for the pyrrolidine molecule, NDC occurs at a bias voltage of less than 0.5 V. The energy of 0.5 eV is of about the same order of magnitude as chemical bond energies, and thus, this NDC was believed to be vibrationally mediated behavior [15]. A scanning tunneling spectroscopy (STS) study using an ultra-high vacuum STM with a Pt-Ir tip showed that the bilayer of a C60 molecule deposited on a Au(111) surface at 7 K has NDC behavior [36] that is multilayer-specific. As such, NDC is not observed when tunneling into a C60 monolayer, a finding that was explained by voltage-dependent changes in tunneling barrier height [36]. In that study [36], the PVR (< 2) could be tuned by adjusting the tip-to-substrate distance. Another way to generate NDC in molecular junctions is to incorporate transition metal atoms in the molecular structure.

A low-temperature (5 K) STM study reported that a cobalt phthalocyanine (CoPc) molecule on a gold substrate exhibited NDC (figure 5) [40]. As shown in figure 5b, NDC was only found to occur for the Ni tip, an outcome that is a result of local orbital symmetry matching between the Ni tip and the Co atom. The NDC effect is reproducible, independent of tip geometrical shape, a finding that contravenes what is known about the mechanism that considers the narrow DOS of the tip apex. Another STM study performed at 12 K under high vacuum conditions by Tu et al. reported the transition (from the presence to the absence of NDC) of copper-phthalocyanine (CuPc) molecules adsorbed on different layers of NaBr grown on a NiAl(110) substrate [18]. The NDC only occurs for one and two atomic layers of NaBr, and there is no NDC for individual CuPc molecules adsorbed on three layers of NaBr. This transition from the presence to the absence of NDC is explained as being due to the opposite bias dependence of the vacuum and NaBr barrier heights and the changing barrier widths for CuPc molecules adsorbed on different layers of NaBr [18]. It is important to note that, for the STM measurements, which locate regularly patterned molecules using molecular resolution images prior to the I-V measurements like those shown in figure 5a, the NDC can be attributed to the individual molecule residing at the site where current signal was measured. On the other hand, for those STM systems where molecules were standing on the substrate surface within a densely packed molecular monolayer instead of lay on the surface, it was still difficult to attribute the measured phenomena to specific molecules.

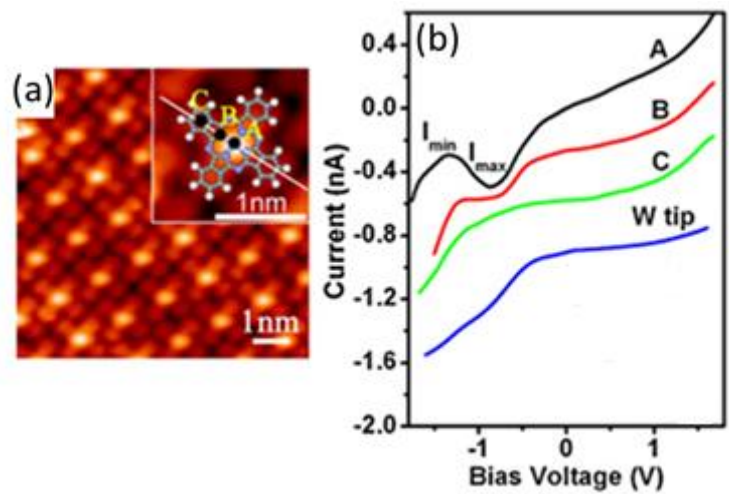

**Figure 5:** (a) CoPc monolayer on a Au(111) surface. The inset is a magnified molecular image. (b) I-V curves measured with a Ni tip over sites A, B, and C (marked in panel a), and with a W tip over site A. Adapted with permission from [40].

As discussed above, the STM based measurements have made great contribution in measuring the NDC behavior of various molecules and understanding the underlying cause. Experimentalists are, however, facing difficulties in determining the exact number of molecules being measured using the STM setup with the tip in touch with target molecules. To put it simply, in this STM system, the NDC behavior was usually attributed to many molecules sandwiched between the tip and substrate, making it hard to recognize the contribution from individual molecules. For those measurements in which the STM tip has no physical contact with molecules, the molecule-tip interface involves a vacuum tunnel distance instead of a solidly-coupled contact through covalent bond between the molecule and tip. In addition, with target molecules adsorbed on only one of the electrodes via covalent bonds and a STM tip positioned above the molecule to be measured by setting a threshold tunneling current that controls the tip-to-substrate distance, it is not easy to perform large number of repeated measurements and the following statistical analysis which have been widely used to discover single-molecule behavior in single-molecule break junction technique.

## 2.3. Measuring weak NDC behavior using the single-molecule break junction technique

Insofar as the nanopore and STM methods are not well-defined, single-molecule-level methods and numerous molecules may be involved in the transport process, neither technique can unambiguously attribute the measured electrical current to a single molecule. Precise measurement of single-molecule conductance is instead facilitated by the single-molecule break junction (SMBJ) technique, which works by repeatedly creating and breaking molecular junctions [42]. It also allows one to measure the current through an individual molecule under a bias sweep, namely, the I-V characteristics. In addition to its capacity to measure single-molecule-level I-V, the SMBJ technique also enables solid molecule-electrode contact via covalent bonds on both sides of the junction, which markedly diminishes the uncertainty, typical in nanopore and STM measurements, at one of the molecule-electrode interfaces. Indeed, the SMBJ technique was recently exploited as a powerful tool to study NDC behavior at the single-molecule level [30, 34, 43-48].

*2.3.1 Measuring NDC – STM break junction (STMBJ) technique:* Tao et al. reported that when sandwiched between Au electrodes at room temperature, the OPE-NO$_2$ molecule (similar to molecule 2 in figure 2, but with both terminals thiolated) exhibited NDC (figure 6) [34]. Although the NDC peaks occur at both positive and negative bias voltages of around ±2 V, their PVRs are markedly different. Typically, the PVR on the high current side of the junction is significantly larger than that on the low current side. Since the asymmetric I-V curves are correlated with the asymmetric location of the nitro moiety, Tao et al. attributed the NDC effect to the electro-active nitro moiety. It was also observed that the NDC peaks usually either decrease or diminish in the reverse voltage sweep, indicating the possible involvement of an irreversible redox process. Interestingly, the magnitude of the PVR measured via the STMBJ technique is comparable to that using STM but much larger than the value obtained using a nanopore technique. More importantly, NDC measured using STMBJ can be unequivocally attributed to the contribution of a single molecular junction with covalent bonds at both molecule-electrode interfaces, which is determined by a statistical analysis over large number of repeated measurements. The NDC effect of OPE-NO$_2$ molecules, therefore, is tremendously enhanced using the STMBJ technique, a finding that is likely due to the better contact coupling at each molecule-electrode interface that, in turn, gives rise to the NDC feature. This finding also highlights the crucial role of the molecule-electrode contact interfaces in the electron transport properties of molecular junctions.

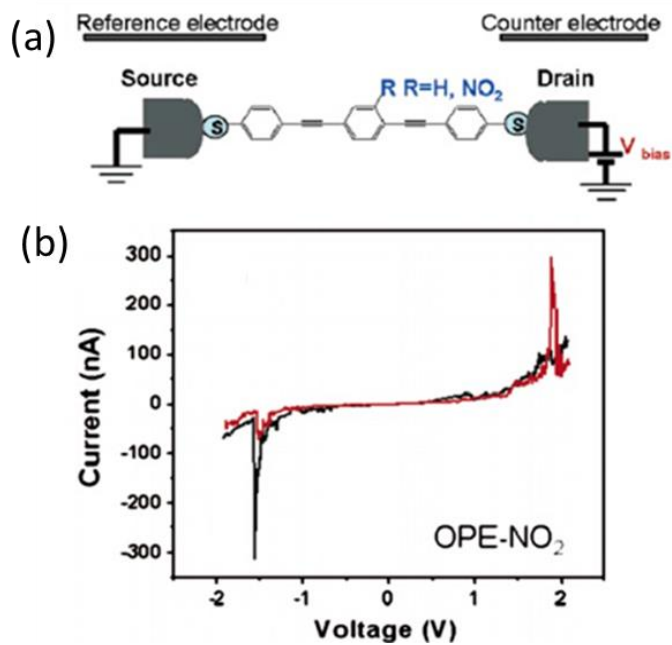

**Figure 6:** (a) Schematic illustration of a single Au-molecule-Au junction. (b) I-V curves of Au-(OPE-NO$_2$)-Au junction. Adapted with permission from [34].

*2.3.2. Measuring NDC – Mechanically controlled break junction (MCBJ) technique*: Recent experimental measurements of a DNA duplex (10-nm long oligomer) showed that it has an NDC effect both in a vacuum and in aqueous solution [30]. Occurring at a relatively high active voltage (2~3.6 V) with a PVR value of around 3 in solution, the NDC behavior of the DNA was attributed to the formation of a polaron, which reorganizes the conducting material at the molecular orbital level (see, e.g., [49]). Additionally, the bias voltages for the NDC peaks were observed to shift to lower biases when the junction was exposed to vacuum conditions, a finding that was explained by the increase in polaron level in the absence of the polarization effects that originate from the water and the ions in solution. Polaron models were also used to explain the NDC behavior of molecules with active redox centers in which electrons can be trapped [30, 50, 51].

The largely observed relatively high active voltage and/or low PVR are central drawbacks to overcome.

Recently, Perrin et al. reported a low bias voltage and a large NDC for a molecule consisting of two conjugated arms connected by a non-conjugated segment under high vacuum at low temperature (6 K) (figure 7) [44]. In their study, the PVR was found to be as large as 15, an unusually high value for a single-molecule device. The mechanism of this NDC is attributed to the intrinsic molecular orbital alignment of the molecule. The low temperature and high vacuum condition may be the major reason of the high PVR as they greatly diminishes the influence of thermo motion and the effects of water and ions in solution, which also makes it perfect for experimental measurement. However, this condition is not applicable for future use of the molecular NDC device. To mimic the daily use of current by a commercial electronic device at room temperature, it is important to explore the low-bias NDC feature of the single-molecule junction under normal conditions (room temperature and normal atmosphere).

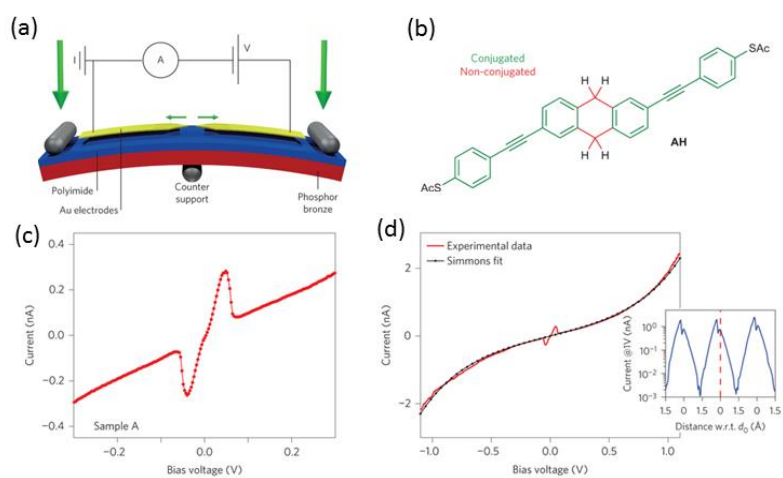

**Figure 7:** (a) Schematics of the experimental MCBJ setup. (b) Structure of a thiolated arylethynylene with a 9,10-dihydroanthracene (AH) core. (c) Typical I-V curves for low bias and (d) the full bias range, the black line is a fit to the Simmons model. Adapted with permission from [44].

*2.3.3. Measuring NDC – conducting atomic force microscopy break junction (CAFMBJ) technique*: Recently, Zhou et al. used the CAFMBJ technique to study the room temperature NDC effect of a molecular junction with a thiol terminated Ru(II) bis-terpyridine (Ru(tpy-SH)$_2$) molecule sandwiched between gold electrodes (figure 8a) [43]. The detailed control of the molecular junctions and the simultaneous measurements of force and conductance revealed new insights into single-molecule NDC behavior. First, NDC behavior is intrinsic to the Ru(tpy-SH)$_2$ molecule. Second, for the Ru(tpy-SH)$_2$ molecule, the NDC only occurs for a specific contact configuration ($G_M$), but not for the other two contact configurations ($G_L$ and $G_H$) (figure 8b). Third, the observed NDR for the specific contact conformation (conductance of $G_M$) is also the result of bias-induced coupling changes (as shown in figure 9), such that the greatest change in force happens at the bias where NDR is observed. The force changes agree with the measurements of electrode interface of single molecule junctions performed by controllable mechanical modulations [52-54], indicating that the force changes are ~~it is~~ likely caused by molecule-electrode coupling changes induced by the bias. It is also possible that the conformational relaxation of molecular junctions is caused by the bias-induced twist of molecular structures, or by bias charging on the redox active center, as in the polaron model [50, 55, 56]. The underlying mechanisms of the conformational changes observed in our system under specific bias conditions, however, still require further systematic study, specifically with respect to theoretical calculations. To the best of our knowledge, Ref. [43] is the only experimental study to report a room temperature, low-bias NDR of a single-molecule junction, such that the NDC occurs at a relatively low bias voltage of around 0.6 V with a PVR of around 1.5. In addition, the room temperature condition implies potentially wider applicability.

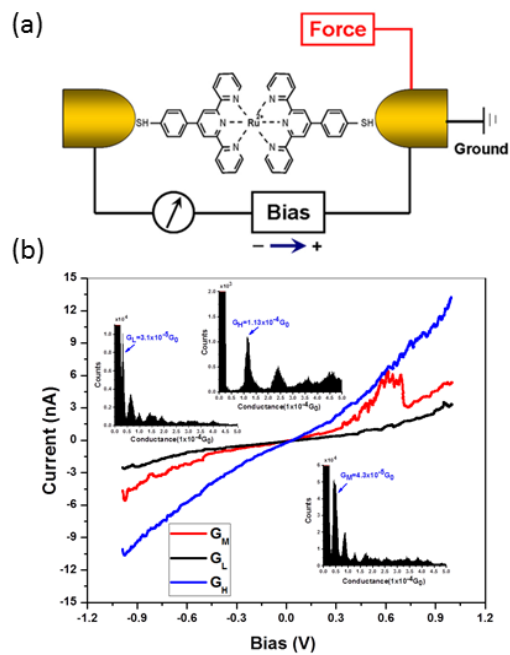

**Figure 8:** (a) Schematic diagram of Ru(tpy-SH)2 molecular junction. (b) Static conductance histograms for three groups under bias voltage of 0.3 V and I-V curves for junctions that lacked the Ru-ion center (black) and for junctions with an Ru-ion center of three sets

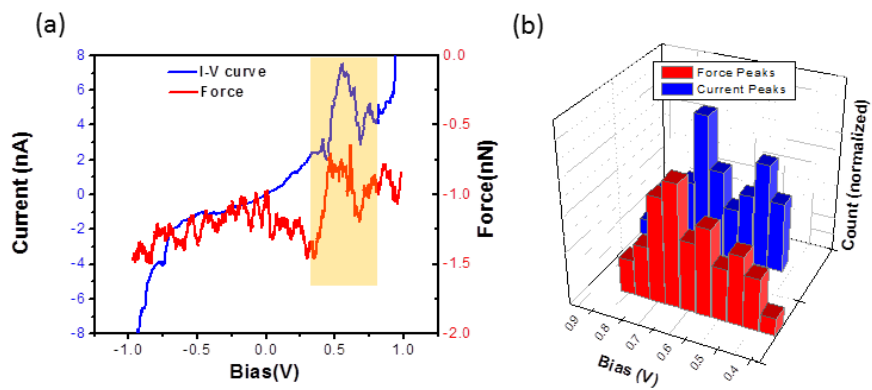

**Figure 9:** (a) Representative monitored I-V curves (blue) and force changes (red) with bias sweeping. The shaded area highlights the NDR and force peaks. (b) The histograms of the NDR peak position (blue) and force peak position (red) using 50 curves each as shown in (a).

### 2.4 Summary of experimental overview

The experimental exploration of molecular NDC behavior – stimulated by the pioneering experiments of Reed et al., who measured a large PVR ratio (1030:1) using the nanopore technique [2, 3, 32, 33] – has undergone significant development in the past fifteen years. Specifically, the capability to determine single-molecule behavior and to better characterize that behavior at the molecule-electrode contact interface – both of which are results of the advent of the single-molecule break junction technique [30, 34, 43, 44] – have greatly advanced the field of NDC measurements. Early NDC measurements, which yielded large PVR ratios, were done using the nanopore and STM techniques at low temperatures under high vacuum conditions [2]. Unfortunately, the same molecule measured at room temperature usually displays a very small PVR ($< 2$) at a relatively high active voltage ($> 2$ V). Considering that a potentially large number of molecules may be involved in the transport process, the contribution made by each individual molecule to that process could be trivial. The development of the single-molecule break junction technique, however, has elucidated the NDC behavior of numerous specific molecular junctions [30, 34, 43, 44]. Although the mechanisms of some NDC behaviors are not fully understood, single-molecule measurements have revealed some of the defining characteristics of NDC, such as low bias and large PVR, which shed light on the nature of molecular NDC. This progress notwithstanding, experimentalists are now confronted with a technical bottleneck as they seek to develop low-bias, room temperature NDC behavior with a more satisfactory PVR value. Meeting this challenge will require both the perfect synthesis of more appropriate candidate molecules and the ability to maintain greater control over molecular junction systems.

## 3. Theoretical approaches to NDC in metal-molecule-metal junctions

In this section we present an overview of the central theoretical explanations and approaches to NDC in molecular junctions. As pointed out in the introduction, this is not meant to be a comprehensive and detailed review, but rather a review of the central mechanisms, demonstrated via simple and physically transparent models, which have been suggested and discussed theoretically to explain NDC.

### 3.1 Calculation approaches

We briefly describe the two principal methods for calculating currents and addressing NDC in molecular junctions (interested readers are referred to one of several recent textbooks devoted to the theory of transport in molecular junctions, e.g., [25, 57]). The *non-equilibrium Green's function* (NEGF), probably the most popular method, was developed in the 1990s to address transport through quantum dots [58], but since then, it has been widely applied in transport through molecular junctions [25, 57]. At its core is the division of the system into three separate regions comprising the left electrode, the right electrode, and the central region (sometimes referred to as the molecular bridge). The electrodes are assumed to be non-interacting and to function as metallic electron reservoirs. Based on this division, the Hamiltonian of the system can be written as

$$\mathcal{H} = \mathcal{H}_L + \mathcal{H}_R + \mathcal{H}_M + \mathcal{H}_{M-L} + \mathcal{H}_{M-R}. \quad (1)$$

where $\mathcal{H}_L$ and $\mathcal{H}_R$ describe the left and right electrodes, respectively, and are typically given by

$$\mathcal{H}_{L,R} = \sum_{k \in \{L,R\}} \epsilon_k c_k^+ c_k + \sum_{k \in \{L,R\}} (V_k c_k^+ d + h.c.), \quad (2)$$

where $c_k$ ($c_k^+$) are the annihilation (creation) operators for electrons in the electrodes.

$\mathcal{H}_M$ describes the central region, which is usually the part that encodes the system's physical characteristics that, in turn, dominate the transport properties. For instance, $\mathcal{H}_M$ can include molecular orbitals, electron-electron interactions, phonons, electron-phonon interactions and more. In many cases (and especially when *ab-initio* calculations are involved), $\mathcal{H}_M$ includes not only the molecular bridge, but also some finite part of the metal electrodes themselves to account for the effects of the molecule-metal interface. In the simplest case, $\mathcal{H}_M$ includes a set of molecular orbitals and is of the form

$$\mathcal{H}_M = \sum_n \epsilon_n d_n^+ d_n. \quad (3)$$

where $d_n$ ($d_n^+$) are annihilation (creation) operators for electrons in a molecular level $|n\rangle$ with energy $\epsilon_n$. Molecule-electrode coupling is then described by

$$\mathcal{H}_{M-L,R} = \sum_{n,k \in \{L,R\}} (V_{k,n} c_k^+ d_n + h.c.). \quad (4)$$

The Hamiltonian of Eqs. 1-4 is the "standard model" for transport through molecular junctions [25, 57]. Once the Hamiltonian is written down, the current $J$ through the junction is given by

$$J = \frac{ie}{2h} \int d\omega \, Tr[(f_L \mathbf{\Gamma}^L - f_R \mathbf{\Gamma}^R)(\mathbf{G}^r - \mathbf{G}^a) + (\mathbf{\Gamma}^L - \mathbf{\Gamma}^R)\mathbf{G}^<] \quad (5)$$

where $e$ is the electron charge, $h$ is Planck's constant, $f_{L/R}$ are the Fermi distribution functions of the left and right electrodes such that $f_{L/R} = \left(1 + \exp\left(\frac{\omega - \mu \mp \frac{V}{2}}{k_B T}\right)\right)^{-1}$ ($\mu$ is the electrode chemical potential, $V$ is the bias, $k_B$ is the Bolzmann constant, and $T$ is the temperature), and $\mathbf{\Gamma}^{L,R}$ are the coupling matrices from the left/right electrode states to the molecular states given by $\mathbf{\Gamma}^{L,R}_{n,m} = 2\pi \sum_k \delta(\epsilon_k - \omega) V_{k,n}^* V_{k,m}$. $\mathbf{G}^{r,a,<}$ are, respectively, the so-called retarded, advanced and lesser Green's functions of the molecule, and they are Fourier-transformed time-ordered correlation functions of the molecule in the presence of electrodes. These can be very hard to calculate, however, and as such, they can be evaluated exactly in only a few limiting cases. If the two electrodes are symmetrically coupled to the molecule, the expression for the current simplifies to

$$J = \frac{e}{h} \int d\omega [(f_L - f_R)] \, Tr(\mathbf{\Gamma}^L \mathbf{G}^r \mathbf{\Gamma}^R \mathbf{G}^a), \quad (6)$$

in which the term $Tr(\mathbf{\Gamma}^L \mathbf{G}^r \mathbf{\Gamma}^R \mathbf{G}^a)$ can be identified with Landauer's transmission function $T(\omega)$, leading to the famous Landauer expression $J = \frac{e}{h} \int d\omega [(f_L - f_R)] T(\omega)$. The external bias enters the calculations through the Fermi functions of the electrodes due to the shifts in their chemical potentials, but this is not enough to fully explain NDC, and it is clear that to obtain a non-trivial effect, the transmission function should also depend on bias. The mechanisms that give rise to such dependence are the focus of this section.

Another method typically used to study transport in molecular junctions entails the rate equations [59-65], an approach that is applicable only when the molecule-bridge coupling is weak (i.e., at the sequential tunneling limit) and at relatively high temperatures. In this approach, the molecular states $|n\rangle$ (in the many-electron Fock space and found from the Hamiltonian of the molecular bridge $\mathcal{H}_M$) are labeled according to their probability of occupation $P_n$. The rate equations can be considered as a classical limit of the more general quantum master equations approach to transport (see, e.g., [66-69], which include not only the probabilities $P_n$, but also coherences between states, represented by the full density matrix. At high temperature, inelastic and environment effects (e.g. dephasing and decoherence) reduce the coherences (off-diagonal elements of the density matrix). The transport properties then depend solely on the diagonal elements of the density matrix, which are the probabilities $P_n$.

The probabilities obey the rate (or master) equations,

$$\dot{P}_n = \sum_m (W_{m\to n} P_m - W_{n\to m} P_m), \qquad (7)$$

where $W_{n\to m}$ is the $(n,m)$ element of the rate matrix $\mathbf{W}$ that describes the rate of transfer from state $|n\rangle$ to state $|m\rangle$. If this $n \to m$ transition is due to electron transfer from the electrodes (i.e., $|n\rangle$ has one electron less than $|m\rangle$), then $W_{n\to m}$ will include the relevant Fermi distributions and will have the form

$$W_{n\to m} = \sum_{\nu=L,R} \gamma_\nu f_\nu(\Delta E_{nm}) \qquad (8)$$

where $\gamma_{L,R}$ is the rate at which the molecule-electrode interface is crossed, and $\Delta E_{nm} = E_m - E_n$ is the difference in energy between the two molecular states. If the $n \to m$ transition is due to electron transfer *to* the electrodes, then the rates will be

$$W_{n\to m} = \sum_{\nu=L,R} \gamma_\nu (1 - f_\nu(\Delta E_{nm})). \qquad (9)$$

Once the rate matrix $\mathbf{W}$ has been determined, the rate equation $\dot{\mathbf{P}} = \mathbf{W} \cdot \mathbf{P}$ can be solved, and the steady-state solution $\mathbf{P}_{ss}$ is the kernel of $\mathbf{W}$. From the definition of the current operator $\hat{J} = e\frac{d\hat{n}}{dt}$ and from the relation between the total charge on the molecule and the probabilities, one can obtain the expression for the current from the steady-state solution $\mathbf{P}_{ss}$. As in the NEGF approach, the bias voltage appears in the Fermi functions of the left and right electrodes, but it may also have a direct effect on the Hamiltonian or the transition rates, as will be discussed below.

In recent years, advances in computing capabilities have facilitated the development of various computational approaches – e.g., density-functional theory (DFT) to generate realistic descriptions of molecular junctions – that complement the above methods and increase their applicability. For example, the most common combination, the so-called NEGF-DFT approach, is used to calculate Green's functions and couplings based on the appropriate Kohn-Sham orbitals (see, e.g., [25, 57, 70-72]). Despite its several important (and possibly critical) flaws [57, 73-75], however, NEGF-DFT has become a standard method in the field of molecular junctions, where in many cases it has generated significant insight into the physical origins of transport phenomena.

### 3.2 Bias-induced changes in molecular orbitals

We begin with what seems to be the simplest and most common explanation for NDC in molecular junctions: bias-induced changes in the molecular orbitals. When a voltage bias $V$ is applied to a molecular junction, there is a voltage drop between the two electrodes. Because the voltage drop occurs where the resistance is largest, one expects most of the drop to occur on the end-groups connecting the molecule to the electrodes, and this is, indeed, typically the case. If the molecule is strongly coupled to the electrodes, however, then it will also experience a substantial drop in its voltage. This will add to the Hamiltonian an additional potential term $\hat{V}(\mathbf{r},V)$ that will depend on the atomic positions and on the applied bias. The bias dependency of the Hamiltonian dictates that Green's functions and the transmission function will also be bias dependent [76, 77].

Using simple examples (below), we show that the typical scenario depicts the transmission function $T(\omega,V)$ (now a function of energy and bias) as *decreasing* with increasing bias. This decrease competes with the increase in the current due to the increase of the integration range (i.e., the Fermi window) in the current formula (Eq. 3), which is a result of the Fermi function difference. If the decrease in the transmission is fast enough to overcome the increase in the Fermi window, the current will decrease while the voltage increases, thus leading to NDC. This scenario has been observed theoretically many times [35, 78-90].

To give an illustrative example, we consider a simple, tight-binding model for a linear chain with $N$ atoms. In a real-space description, the system is modeled by the non-interacting spinless Hamiltonian connected at sites $i = 1$ and $i = N$ to external electrodes,

$$\mathcal{H}_M = \epsilon_0 \sum_{i=1}^{N} d_i^+ d_i - t \sum_{i=1}^{N-1} d_i^+ d_{i+1} + h.c.,$$

$$\mathcal{H}_{M-L,R} = \sum_{k \in L}(V_{k,1} c_k^+ d_1 + h.c.) + \sum_{k \in R}(V_{k,N} c_k^+ d_N + h.c.) \qquad (10)$$

where $d_i$ ($d_i^+$) annihilates (creates) an electron on the atom at position $i$.

The electrodes are assumed to have a constant density of states (DOS) and a resulting energy-independent constant, imaginary only, known as self-energy (the so-called wide-band approximation [91]). The voltage drop across the molecule is characterized by a parameter $0 \leq \alpha \leq 1$ that sets the ratio between the total voltage drop $V$ and the actual voltage drop on the molecule. With this parameter, an additional term is added to the Hamiltonian of the form

$$\mathcal{H}_b = \alpha V \sum_{i=1}^{N} \left(\frac{n-1}{N-1} - \frac{1}{2}\right) d_i^+ d_i, \quad (11)$$

which guarantees that the potential is $-\alpha \frac{V}{2}$ at the leftmost site and $\alpha \frac{V}{2}$ at the right most site, with an overall voltage drop of $\alpha V$ across the molecule.

We start with the simplest possible example, a single molecular orbital (i.e., $N = 1$). In this case the transmission function is a shifted Lorenzian,

$$T(\omega, V_{bias}) = \frac{\Gamma^2}{\Gamma^2 + (\omega - \epsilon_0 + \alpha V)^2}, \quad (12)$$

where $\Gamma$ is the electrode-induced level broadening (which we assume is symmetric). Eq. 12 already demonstrates the mechanism responsible for NDC: as the voltage increases, the effective position of the molecular resonance is shifted away from the Fermi level, a process which competes with the increase of the Fermi window.

The Fermi functions at zero temperature become step-functions, and the integral can be evaluated analytically, yielding

$$I = \frac{e\Gamma}{h}\left(\operatorname{atan}\left(\frac{\epsilon_0 + \left(\alpha - \frac{1}{2}\right)V}{\Gamma}\right) - \operatorname{atan}\left(\frac{\epsilon_0 - \left(\alpha + \frac{1}{2}\right)V}{\Gamma}\right)\right). \quad (13)$$

NDC will occur if $I(V)$ has a maximum in $V$. Simple algebra reveals that $I(V)$ has a maximum at $V_{mx} = \pm 2\sqrt{\frac{\epsilon_0^2 + \Gamma^2}{4\alpha^2 - 1}}$, implying that NDC occurs only for $> \frac{1}{2}$. That is, when the voltage drop on the molecule is large enough, the position of the energy level shifts with $V$ to compensate for the Fermi window opened by the integration over the Fermi functions, and the current can decrease with voltage.

However, this is neither a generic nor a universal feature. In fact, already for $N = 2$ (for which the current can also be calculated analytically, albeit via more tedious algebra) one finds that there is always a maximum for $I(V)$. In figure 10, the current-voltage ($I - V$) curve is plotted for an $N = 2$ molecule for different values of $\alpha$ ($\alpha = 0, 0.25, 0.5, 0.75, 1$). Other numerical parameters are $\Gamma = 0.01$ eV, $t = 0.1$ eV and $\epsilon_0 = 0.3$. The top panels of figure 10 show a schematic of the model at zero bias (top left) and at finite bias (top right). The central characteristic of the model is the sharpening of the NDC feature as $\alpha$ increases, i.e., an increase in the voltage drop across the junction.

The central drawback of this simple phenomenological model for NDC is its failure to correctly take into account the voltage drop. In realistic systems, the voltage drop is associated with the electron density $\rho(\mathbf{r})$ across the molecule through the Poisson equation $\nabla^2 V(\mathbf{r}) = -\rho(\mathbf{r})/\varepsilon(\mathbf{r})$ (where $\varepsilon(\mathbf{r})$ is the permittivity). Therefore, one must calculate self-consistently the density (which can be evaluated with the use of the lesser Green's function, for instance, or using the rate equations) and the potential drop. This type of self-consistent voltage calculation has been implemented in various DFT-based transport calculations [70-72, 92, 93].

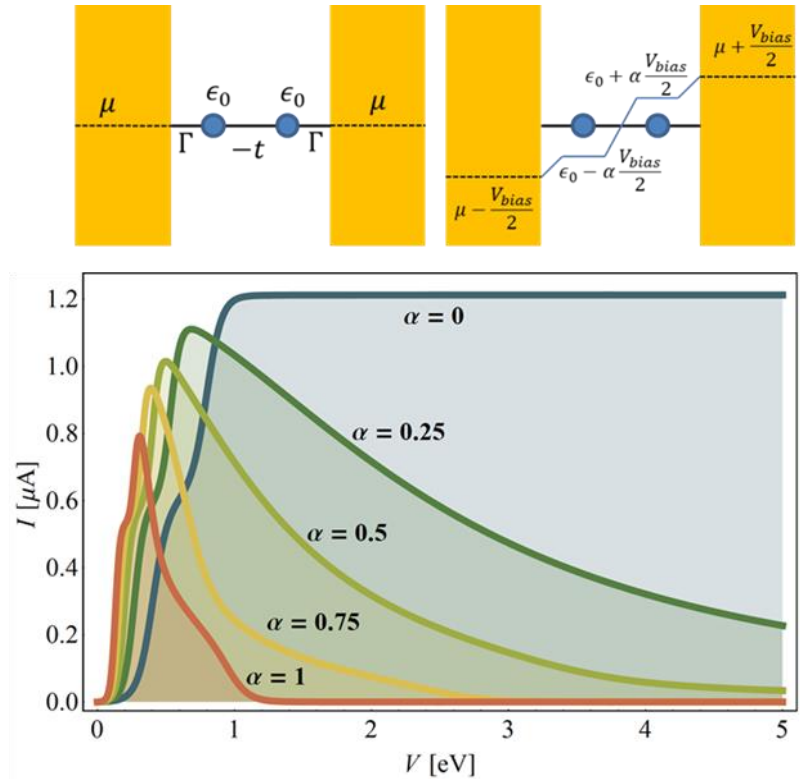

**Figure 10:** Top: schematic depiction of a molecular junction of length $N = 2$ at zero bias (top left) and finite bas (top right). Bottom: $I - V$ characteristics for differet values of $\alpha$, ($\alpha = 0, 0.25, 0.5, 0.75, 1$), representing in increasing voltage drop across the junction. The NDR features sharpen as the voltage drop on the molecule increases.

To demonstrate here the role of self-consistency between the potential and density, we follow the approach of Mujica, Roitberg and Ratner [77], who discretized the Poisson equation for the simple-tight-binding molecular model. The resulting equation for the voltage drop reads

$$\frac{1}{a^2}(V_{i+1} - 2V_i + V_{i-1}) = -\frac{\delta\rho_i}{\varepsilon}, \tag{14}$$

where $a$ is the inter-atomic distance and $\delta\rho_i$ is the deviation of the density on the $i$-th site from the equilibrium, zero-bias state. We consider a molecule of length $N = 2$, with $\Gamma = 0.01$ eV, $\epsilon_0 = 0.3$ eV and $t = 0.2$ eV. The local electronic density $\rho_i$ ($i = 1,2$) is evaluated using the lesser Green's function according to $\rho_i = \int \frac{d\omega}{2\pi} \text{Im} G_{i,i}^<(\omega)$, where Green's function is evaluated self-consistently with Eq. 14. The $I - V$ characteristics are plotted in figure 11 for temperatures $T = 300$ K (left) and $T = 60$ K (right). Two features are readily apparent: the first is the NDC, which occurs as in figure 10, due to the change in the position of the electronic resonance with the voltage drop. A second feature, apparent only at low temperature, is the presence of *hysteresis*, that is, differences in the $I - V$ curves when scanning toward the positive bias (blue curve) vs. when scanning toward the negative bias (orange curve) (scanning direction indicated with arrows). Hysteresis implies bistability, which is due to the non-linear relation between the potential drop and the density.

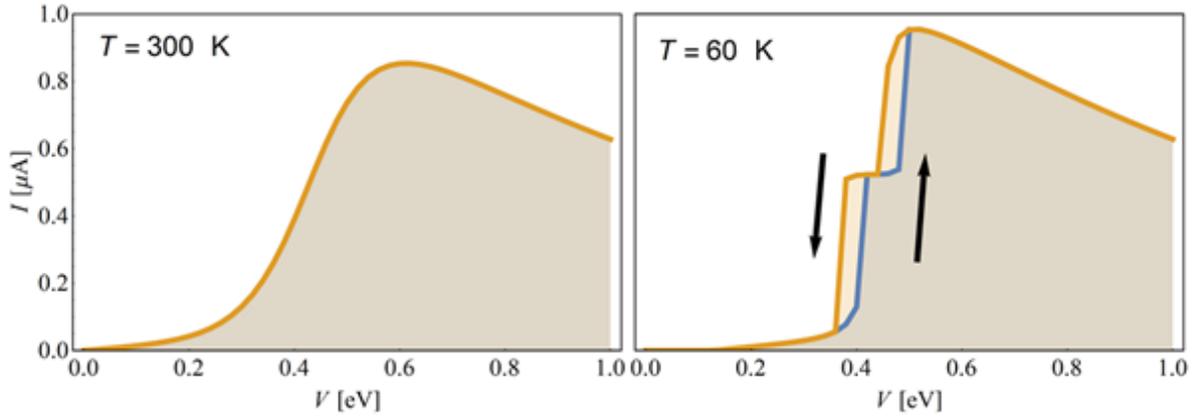

Figure 11: $I - V$ curves for a $N = 2$ molecular junction with a voltage drop determined self-consistently with the density (Eq. 14) for temperatures $T = 300$ K (left) and $T = 60$ K (right). In addition to NDC, there is hysteresis due to density bistability arising from the non-linear relation between the bias and the density.

The bias can induce changes not only in the molecular orbital, but also in the coupling between the molecular levels and the electrode, which can also lead to NDC. To simply demonstrate this outcome, we again look at the Hamiltonian of a simple chain, Eq. 10. However, here we begin by diagonalizing the molecular Hamiltonian $\mathcal{H}_M$, which now takes the diagonal form $\mathcal{H}_M = \sum_n \epsilon_n d_n^+ d_n$. The molecule-electrode part of the Hamiltonian is now given by

$$\mathcal{H}_{M-L,R} = \sum_n \sum_{k \in L} \left( V_{k,1}^{(n)} c_k^+ d_n + h.c. \right) + \sum_{k \in R} \left( V_{k,N}^{(n)} c_k^+ d_n + h.c. \right) \tag{15}$$

where now $V_{k,1}^{(n)} = V_{k,1} \psi_n(1)$ and $V_{k,N}^{(n)} = V_{k,1} \psi_n(N)$, in which $\psi_n(i)$ comprises the wave-functions that diagonalize the molecular Hamiltonian. This implies that level-broadenings (i.e., the imaginary-parts of the self-energies of the electrodes) from the left and right electrodes will (i) become non-symmetric, and (ii) be normalized by $|\psi_n(1)|^2$ and $|\psi_n(N)|^2$, respectively. When voltage drops on the molecule (as in Eq. 11), it will affect the wave-functions, which, in turn, subsequently affects the self-energies and the transmission function.

In figure 12a, the square of the wave-function $|\psi_n(i)|^2$ is plotted as a function of position for a chain of length $N = 8$ (other parameters are the same as for figure 10) for different voltage bias values ($V = 0, 0.5, 1, 1.5, 2$; we set $\alpha = 0.4$). We choose the $n = 4$ molecular level, which is the HOMO level for a half-filled molecule, but the results are similar for all states. As seen, due to the voltage drop, the wave-functions develop a distinct left-right asymmetry and show a strong (exponential) decrease of the weight of the wave-function on the left- and right-most edge sites (note the logarithmic scale). In figure 12b, the orbital weight on the left-most site ($|\psi_L|^2 = |\psi_{HOMO}(1)|^2$) is plotted as a function of voltage drop for different molecular lengths.

This change in the molecular orbitals implies that the level broadening, even in the wide-band approximation, decreases exponentially with voltage bias. The $I-V$ curves of a molecule with length $N=8$ (all parameters are the same as in figure 12a) for two cases show the effect of this decreased coupling on current (figure 13). In the first case (dashed blue line), the self-energy is kept constant (as in figure 10). In the second case, the self-energies are renormalized according to $\Gamma_{L,R} = \Gamma |\psi_{L,R}|^2$, in accordance with the arguments above. As can be seen, taking into account the bias-induced change in the molecule-electrode coupling leads to NDC. The explanation is straight-forward: the width of the transmission bias is proportional to $\Gamma$, and a broader transmission yields larger currents. Therefore, the reduction of $\Gamma$ with bias competes with the increase of the Fermi window, leading to a maximum in the current and to NDC.

This is, of course, a simplified picture of the dependence of coupling on bias voltage. This mechanism was recently revealed as a possible origin of NDC, where the presence of both the voltage bias-induced field effect along the molecule and Coulomb interactions (considered within a DFT calculation) showed that mixing between various orbitals induces changes in the coupling between the orbitals and the electrodes [94].

### 3.3 Electron-phonon interactions

Galperin *et al.* [49, 95] suggested a different mechanism for NDC based on electron-phonon coupling. The general idea is that the electronic orbital energy (and possibly the coupling between the molecular orbitals and the electrode states) renormalize due to the electron-phonon coupling. This renormalization depends on the electron occupation and, therefore, on bias voltage. In certain parameter ranges, the renormalization is such that it reduces the transmission function with increasing bias, and the end result is NDC.

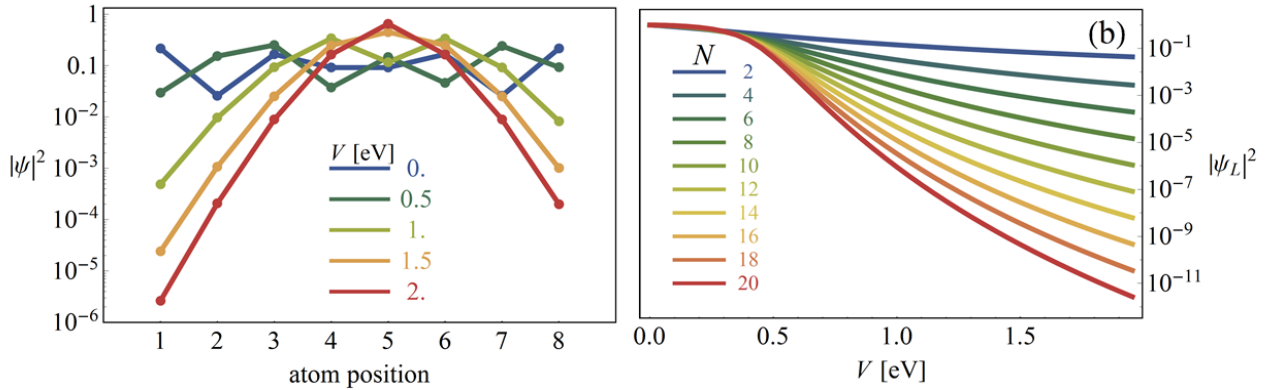

Figure 12: (a) Square of the HOMO wave-function as a function of position for a chain of length N=8 (other parameters are the same as for figure 3.1) for different voltage bias values ($V = 0, 0.5, 1, 1.5, 2$; and $\alpha = 0.4$). An exponential decrease of the orbital weight in the edge sites is clearly seen. (b) Orbital weight on the left-most site ($|\psi_L|^2 = |\psi_{HOMO}(1)|^2$) as a function of voltage drop for different molecular lengths.

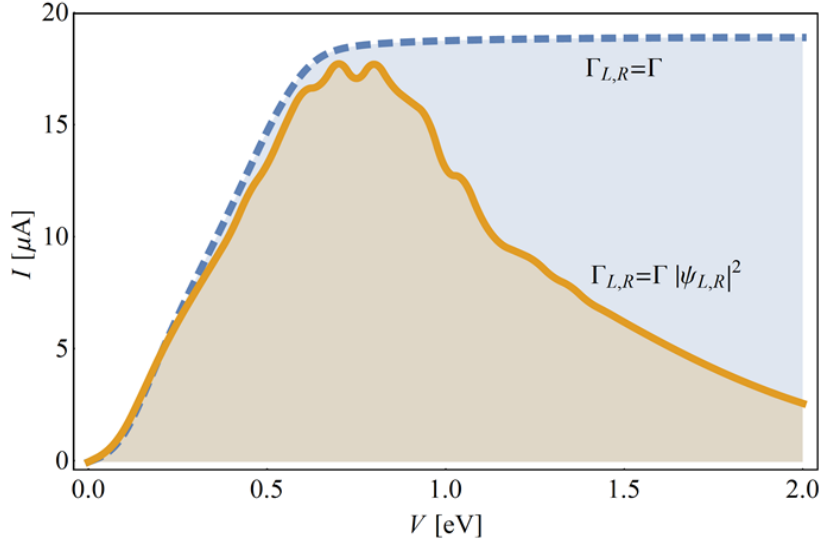

**Figure 13:** the $I - V$ curves of a molecule with length $N = 8$ (other parameters are the same as in figure 12a) for a junction with a constant self-energy (dashed blue line), and self-energies are renormalized according to $\Gamma_{L,R} = \Gamma \left|\psi_{L,R}\right|^2$ (orange solid line). The reduction of $\Gamma_{L,R}$ due to the bias voltage induces NDR (see text).

The starting point of the calculations by Galperin *et al.* is a Hamiltonian for the molecular junction, including electrons and phonons, of the form:

$$\mathcal{H} = \mathcal{H}_e + \mathcal{H}_{ph} + \mathcal{H}_{e-ph}$$

$$\mathcal{H}_e = \epsilon_0 d^+ d + \sum_{k\in\{L,R\}} \epsilon_k c_k^+ c_k + \sum_{k\in\{L,R\}} (V_k c_k^+ d + h.c.)$$

$$\mathcal{H}_{ph} = \omega_0 a^+ a + \sum_q (\omega_q b_q^+ b_q + U_q(a^+ + a)(b_q^+ + b_q))$$

$$\mathcal{H}_{e-ph} = \lambda(a^+ + a) d^+ d$$

(16)

where $d, c_k, a$ and $b_q$ (and their adjoints) are annihilation (creation) operators for electrons on the molecule, electrons in the electrodes, a phonon on the molecule (so-called primary phonon) and bath vibrations, respectively. $\epsilon_0, \epsilon_k$ are electronic energies in the molecule and electrode, respectively, and $\omega_0, \omega_q$ are vibrational frequencies for the molecular phonon and bath phonons respectively. $V_k$ is the molecule-electrode coupling, $U_q$ is the molecular phonon-bath phonon coupling, and $\lambda$ is the molecular electron-phonon coupling. This is the "standard model" in the study of the vibrational effects on transport through molecular junctions [25, 96-107].

The authors then employ a series of approximations, including a Born-Oppenheimer approximation, which essentially decouples the electron and phonon dynamics (due to the very different time-scales involved in the motions of these two particles), a wide-band approximation for the phonon bath and electronic states in the electrodes, and a mean-field approximation for the phonon operators. These approximations, described in greater mathematical rigor in [80], lead to a very simple renormalization of the molecular level,

$$\tilde{\epsilon}_0(\rho) = \epsilon_0 - \frac{2\lambda^2 \omega_0}{2\omega_0^2 + \left(\frac{\gamma}{2}\right)^2} \rho \qquad (17)$$

where $\rho$ is the electron density on the molecule, and it is, as described in section 3.2, bias dependent.

The calculation of the density (evaluated from the lesser Greens' function in the presence of bias) and the renormalized molecular energy should be performed self-consistently. As in the case of the Poisson equation, this self-consistent calculation can have multiple solutions, pointing to a bistability and a resulting NDC and hysteresis. Figure 14, reproduced with permission from Ref. [49], shows the $I - V$ curve, where the NDC is visible, and strongly resembles the experimental results by Tour et al. [2] displayed in figure 3.

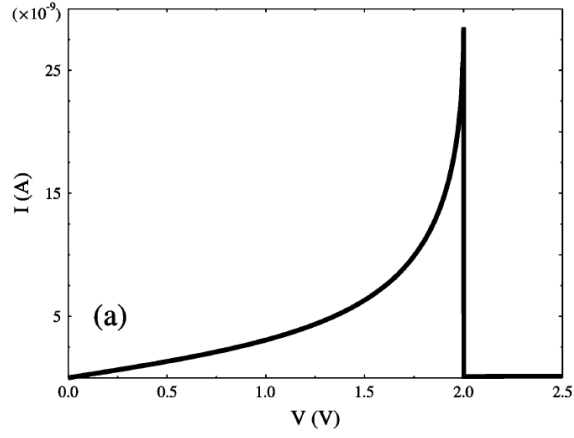

Figure 14: I-V curve of the model presented in Eqs. 12-13, a molecular junction in the presence of vibrations, to be compared with figure 3. Adapted with permission from Ref. [49].

More insight into the origins of NDC as possibly from electron-phonon coupling can be obtained by looking at the so-called Lang-Frisov polaron transformation for the Hamiltonian of Eq. 16, which for simplicity lacks the interaction with bath phonons (i.e., $U_q = 0$) [108, 109]. For this aim, one defines the operator $S = \exp[\gamma\lambda(a^+ - a)d^+d]$. The transformed Hamiltonian $\tilde{\mathcal{H}} = S^+\mathcal{H}S$ now reads

$$\tilde{\mathcal{H}} = \tilde{\epsilon}_0 d^+ d + \omega_0 a^+ a + \sum_{k\in\{L,R\}} \epsilon_k c_k^+ c_k + \sum_{k\in\{L,R\}} (\tilde{V}_k c_k^+ d + h.c.), \qquad (18)$$

where $\tilde{\epsilon}_0 = \epsilon_0 - \lambda^2\omega_0\gamma(2-\gamma) + \lambda\omega_0(1-\gamma)(a^+ + a)$ and $\tilde{V}_k = V_k e^{-\lambda\gamma(a^+-a)}$. In the absence of electrodes ($V_k = 0$) the solution $\gamma = 1$ fully eliminates the electron-phonon interaction, but when electrodes are present, $\gamma$ is non-zero, and its optimal value can be evaluated using a Monte-Carlo variational calculation. For finite $\gamma$, it is clear that the local energy will depend on phonon-occupation that, in turn, will depend on the electron occupation. Similar to the case presented above, voltage bias will then induce a change in the molecular occupation that will cause a shift of the effective molecular orbitals, thereby resulting in NDC. Recent investigations of this effect, also with the presence of electron-electron interactions in the molecule, revealed a delicate balance between Coulomb interactions and polaron formation [56, 110-113].

### 3.4 Coulomb effects

Thus far, the theoretical discussion has been limited to non-interacting electrons, and the Coulomb repulsion between electrons was discarded. However, the Coulomb repulsion may be a dominant factor in molecular junctions, leading to effects such as the experimentally observed Coulomb blockade [114, 115]. As discussed below, Coulomb interactions can also lead to NDC via several possible mechanisms.

*3.4.1 Coulomb Blocking*: NDC due to Coulomb interactions – and specifically, Coulomb blocking – was suggested by Muralidharan & Datta [60]. Consider a molecular junction, generically modeled by its HOMO and LUMO levels. The Fock space description of the molecular junction will then include four states, $|\emptyset\rangle, |1\rangle, |2\rangle$ and $|D\rangle$, corresponding respectively to an empty molecule (i.e., an electron removed from the HOMO), a molecule singly occupied by an electron in the HOMO, a molecule singly occupied by an excited electron in the LUMO, and a doubly-occupied molecule whose electrons occupy both the HOMO and the LUMO. The total energy of these states is then $E_\emptyset = 0, \epsilon_1, \epsilon_2$ and $E_D = \epsilon_1 + \epsilon_2 + U$, respectively. The Coulomb energy $U$ is the shift in the LUMO energy due to the presence of an electron in the HOMO, the result of the Coulomb repulsion between the two electrons. The energy difference $\Delta E_{opt} = \epsilon_2 - \epsilon_1$ is referred to as the optical gap, $\Delta E_f = E_D - \epsilon_1$ is referred to as the "fundamental gap" or the "transport gap", and the difference between the two gaps is the so-called exciton binding energy [116]. This model is depicted in figure 15, where the HOMO and LUMO levels of a bipyridyl-dinitro oligophenylene ethynylene (PBDM) molecule are shown, as calculated by DFT (adapted with permission from [117]). These orbitals are localized on the molecular bridge and couple differently to the electrodes, leading to the effective two-level model described above and schematically depicted in figure 15.

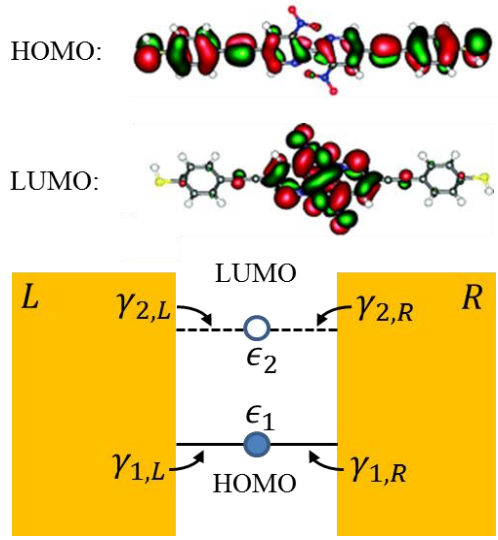

**Figure 15: HOMO and LUMO of PBDM molecule (adapted from Ref. [117] and the effective two-level model. Due to the weak coupling between the LUMO and the electrodes, it plays the role of a "blocking state", which leads to NDC (see text).**

Considering the case in which the Coulomb energy $U$ is large, the doubly-occupied state is practically unimportant, as its energy is much higher than that of the LUMO level, and the molecular junction can be practically described with three levels. In the simplified picture of weak coupling (Eqs. 7-9), the coupling to the left (L) and right (R) electrodes is described by the transfer rates of electrons from the HOMO and LUMO levels to the L and R electrodes, $\gamma_{\nu,X}, \nu = 1,2, X = L,R$. Let us first discuss the case of equal transfer rates (depicted in figure 15) with the numerical parameters $\gamma_{1,L} = \gamma_{1,R} = \gamma_{2,L} = \gamma_{2,R} = 10^{-3}$eV, $\epsilon_1 = -0.1$eV, $\epsilon_2 = 0.5$eV, $T = 300K$ and $U = 8$ eV.

At zero bias, the HOMO level may or may not be occupied, depending on whether it is below or above the electrode's Fermi level (here we describe only the first case, i.e., below electrode Fermi level, but the process is similar for both cases), and the LUMO level is unoccupied. Once a voltage bias is applied, the Fermi window opens, and the LUMO level begins to fill as the left electrode's chemical potential reaches it. Since the two levels cannot be occupied simultaneously (because of the Coulomb charging energy), the increased occupation of the LUMO induces a decrease in the HOMO occupation (see figure 16a). However, because the HOMO and LUMO are coupled equally well to the electrodes, they become equivalently populated.

To relate occupation to transport, we show that transport through the levels occurs when an electron hops onto a level from one electrode and then off that level to the other electrode. Since only one electron can occupy the level at any given time, on average the current will be proportional to the occupation of the level (or specifically, to the difference between the occupation at finite bias and the occupation at equilibrium),

$$I_1 = \gamma_{1,L}\left(n_1 - n_1(V=0)\right), I_2 = \gamma_{2,L}\left(n_2 - n_2(V=0)\right),$$
$$I_{\text{tot}} = I_1 + I_2 \ . \qquad (19)$$

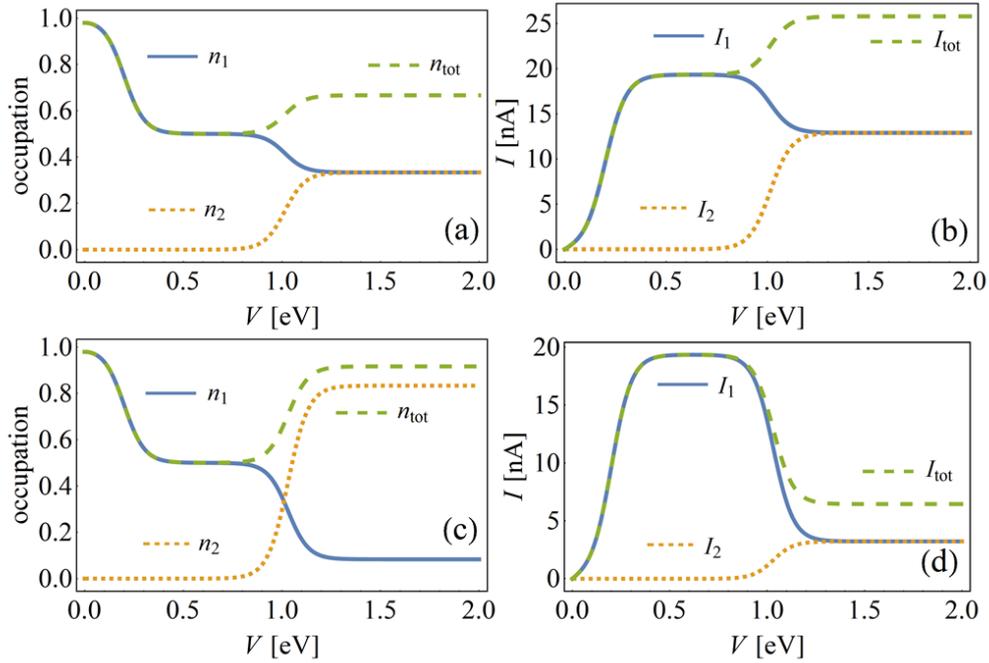

Figure 16: (a) Occupation of the HOMO level $n_1$, LUMO level $n_2$ and total occupation $n_{tot} = n_1 + n_2$ as a function of bias for the fully symmetrical molecular junction. (b) Current through the LUMO ($I_1$), the LUMO ($I_2$) and total current for the symmetrical molecular junction, showing no NDC. (c-d) same as (a-b), for a molecular junction where the LUMO is asymmetrically coupled to the electrodes (see text for parameters).

Therefore, although the current through the HOMO level decreases (figure 16b), it is compensated by the increase in current through the LUMO, such that the total current increases. This general feature is apparent when the couplings between the levels and the electrodes are of similar magnitude.

However, the situation is completely different if the LUMO level is only weakly coupled to the drain electrode (the electrode with the lower voltage; the right electrode in this example). In this case, once the voltage reaches the LUMO level, it again begins to fill. However, because it is now harder for electrons to leave the LUMO level (due to its weak coupling to the drain electrode) and because double-occupation is still forbidden due to the Coulomb repulsion, the HOMO level begins to empty of its electrons at a faster rate than the LUMO level, resulting in depletion of the former level. This can be seen in figure 16c, where the same calculation is performed as in figure 16a, with the only change being $\gamma_{2,R} = 10^{-4}$ eV $= 0.1\, \gamma_{1,R}$. In this scenario, the excited level is called a "blocking state", the resulting current through which is not enough to compensate for the reduction in current due to HOMO level depletion, and therefore, the overall current *decreases* with increasing current (figure 16d), leading to NDC behavior.

The authors of Ref. [60] present very general conditions for the occurrence of Coulomb-blocking induced NDC, which depend on the direction of the bias. In the case described above, for instance, the condition is the somewhat intuitive $\frac{1}{\gamma_{2,R}} > \frac{1}{\gamma_{1,L}} + \frac{1}{\gamma_{1,R}}$. Recently, this mechanism for NDC was generalized to any situation in which there are two conduction channels, such that the conduction of one channel depends on the occupation of the other (as in the case of the blocking state). The oxidation states in so-called redox molecular bridges are one example, and the roles of fluctuations and reorganization were discussed in detail [28, 117-120]. Interestingly, this phenomenon was also observed in other nano-scale junctions, for instance, STM-based junctions with metallic shell structures [121].

*3.4.2 Coulomb interactions with electrode electrons*: Another Coulomb-interaction mechanism that can lead to NDC comprises the interactions between the molecular bridge electrons and electrons on the electrodes. Such molecule-electrode Coulomb interactions may induce an asymmetry (that then leads to NDC via the mechanism described above), renormalize the molecular levels, or change the molecule-electrode couplings [94, 122-125].

It is first important to note that in many calculations, all interaction effects related to the electrodes are typically neglected, and the electrode electrons are considered to be non-interacting [25, 57, 58]. This approximation is typically justified by the claim that interactions in the electrodes are screened, but in molecular junctions and close to the interface, this may not be the case [126].

Simply put, interface Coulomb interactions can be understood as an electrostatic image-charge effect. When a point charge is placed between two parallel plates, it feels an electrostatic potential due to the formation of image-charges in the electrodes, which are nothing but Coulomb-induced redistributions of the charges in the metal. In the limit of infinite perfect plates separated a distance $d$ from each other and a point charge placed at distance $x$ from one plate, the image charge potential is

$$V_{im}(x) = 0.36\, n\left(\frac{1}{x} + \frac{1}{d}\left(H_0\left(\frac{x}{d}\right) + H_0\left(-\frac{x}{d}\right)\right)\right) \text{eV},$$

where $n$ is the charge on the molecule, $H_0(\xi)$ is the Euler's harmonic number, and $d$ and $x$ are in nanometers. Assuming that the central charge is between the plates (i.e., $x \sim d/2$) yields $V_{im} \sim \frac{n}{d}$. This potential shifts the position of the molecular orbirtals, leading to a density-dependent transmission function as in Eq. 12 and resulting in NDC. In longer molecules, a similar situation may develop, with the difference that the image charge potential is not uniform along the molecule. This, in turn, will affect not only the energy but the orbitals themselves – and consequently, the coupling between the orbitals and the electrode states – again resulting in NDC. Finally, a local shift in orbital energy can also arise from the Coulomb interaction between the electrons on the molecule and those in the surrounding solvent [127], with the end result very similar NDC behavior to that experimentally observed (figure 3).

*3.4.3 Excitonic coupling across the interface*: Recently, a different mechanism for weak NDC based on a combination of excitonic interaction (as in Coulomb blocking) and interface Coulomb effects was suggested [128]. In this scenario, when an electron hops from the bridging site to the molecular bridge (say, the LUMO), it leaves a hole behind. The Coulomb attraction between this hole and the electron on the molecule bind them into an electron-hole pair (i.e., exciton) bound to the molecule-electrode interface (depicted in figure 17). The presence of the bound exciton renormalizes the molecule-electrode coupling and leads to NDC.

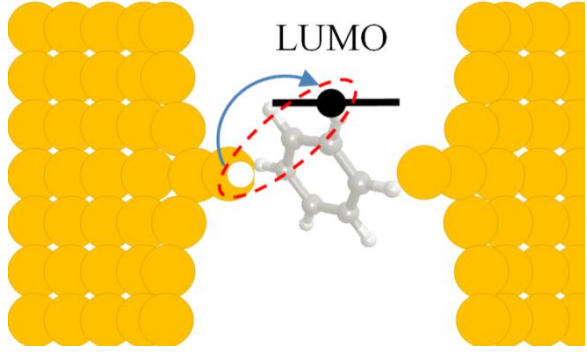

**Figure 17: Schematic representation of the formation of a bound exciton in the molecule-electrode interface (adapted from Ref. [128].**

To model this, we use the standard molecular Hamiltonian of Eqs. 1-4, together with an additional interaction term

$$\mathcal{H}_I = U_1(\hat{n}_L \hat{n}_d + \hat{n}_R \hat{n}_d) \qquad (20)$$

where $\hat{n}_d$ is the density operator on the molecule, and $\hat{n}_{L,R}$ are density operators for electrons on the left and right electrodes, specifically on the atoms closest to the molecules (the so-called bridge atoms, figure 17). A mean-field decoupling of $\mathcal{H}_I$ results in an effective coupling between the molecule and the electrode levels of the form

$$\mathcal{H}_T = -\tilde{t} \sum_\sigma (c^+_{\sigma,L} d_\sigma + c^+_{\sigma,R} d_\sigma + \text{h.c.}) \qquad (21)$$

where $\sigma$ is the electron spin, and $c^+_{\sigma,L/R}$ creates an electron on the bridge atoms closest to the molecule on the left/right electrodes. Assuming a constant DOS $\rho_0$ in the electrodes (the wide-band approximation), the effective level broadening that appears in Eq. 5 becomes $\Gamma \approx \rho_0 \tilde{t}^2$, where $\tilde{t} = t + \tau$, where $\tau = U_1 \sum_\sigma \langle c^+_{L,\sigma} d_\sigma \rangle$ is the exciton amplitude. Calculable within the NEGF formalism, $\tau$ depends on the Fermi distributions of the left and right electrodes and, therefore, on bias. Consequently, the effective molecule-electrode coupling becomes bias-dependent. In figure 18, the coupling (red) and current (blue) are plotted as a function of bias. The coupling is found to be non-monotonic, with a maximum close to the bare transmission resonance. Consequently, the current shows NDC behavior. A comparison of these results with figure 9a shows excellent qualitative agreement. Furthermore, this understanding can explain why the NDC is evident only for intermediate couplings, as observed in Ref. [43], which is due to two competing processes, namely, the reduction of the interface Coulomb repulsion and reduction of the bare molecule-electrode coupling for increasing molecule-electrode distances.

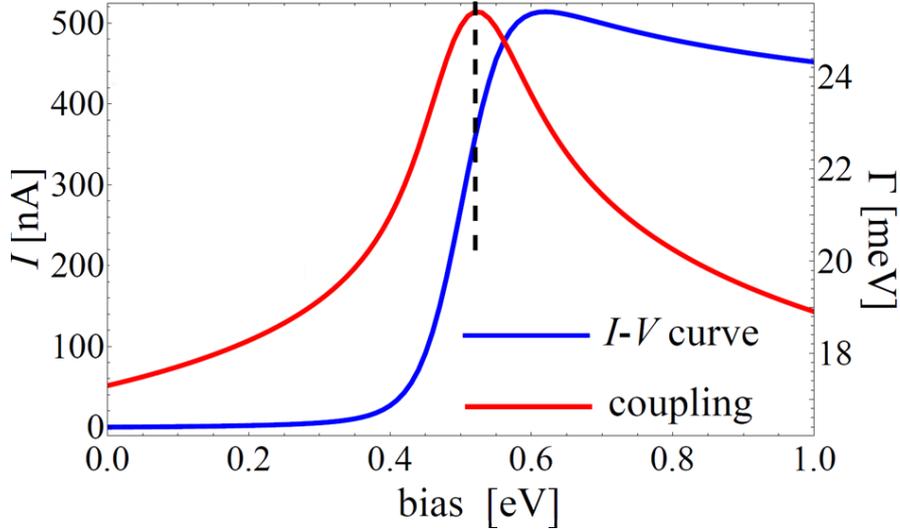

Figure 18: Molecule-electrode coupling Γ (red) and current as a function of bias voltage in the presence of exciton binding across the interface. NDC is showing due to an inhomogeneous dependence of Γ on bias. This should be qualitatively compared with the shaded area in figure 9a. Adapted with permission from Ref. [128].

### 3.5 Summary of theory overview

Rather than serve as an "inventory" of past studies, this section exploits the available theories to provide a clear picture of the possible mechanisms of NDC. We divided the possible mechanisms into three groups based on NDC induction, namely, from the bias voltage drop on the molecule, from interactions between electrons and phonons in the molecular bridge, and from various forms of Coulomb interaction, either in the molecular bridge or between the electrons on the molecule and those on the electrodes. While all three mechanisms reflect changes in the molecular junction, other possibilities, such as the narrowing of the density of states of the tip apex in STM measurements, have also been suggested [7, 36, 90].

It is possible that actual experiments will show that more than one of these mechanisms is at work simultaneously in any given scenario. However, to improve the designs of future molecular junctions exhibiting NDC by incorporating higher PVRs and lower NDC voltage bias onsets, the dominant NDC mechanism in any given experimental setup must be fully understood. To that end, in Table 1 we summarized two important experimental findings vis-à-vis NDC – namely, whether hysteresis is present and the degree of temperature dependence of the NDC – for different mechanisms. However, this list is partial, as temperature dependence and the appearance of hysteresis were not investigated for all the NDC mechanisms. These features can be easily tested experimentally to distinguish between the different mechanisms, and as such, they can be used to interpret future experimental results.

| mechanism | Temperature dependence | Hysteresis |
|---|---|---|
| Voltage drop along the molecule | Strong (figure 11) | Yes (figure 11) |
| Electron-phonon coupling | Strong [49] | Yes [49] |
| Coulomb blocking | Weak | Yes [117] |
| Exciton coupling | Weak [128] | None [128] |

Table 1: The extent of temperature dependence and the presence (or absence) of hysteresis in NDC based on the theoretical descriptions of the different mechanisms for NDC.

### 4. Concluding remarks

The plethora of both theoretical and experimental works in the field of NDC has generated vast amounts of interesting, useful and even controversial information about the NDC phenomenon in metal-molecule-metal junctions. In this overview, we sought to clarify and simplify the overall theoretical and experimental picture of NDC in metal-molecule-metal junctions. It is clear that the road to definitively clarifying the origin of NDC via experiments – a necessary step in the design of future NDC molecular devices – is still very long, but the collaboration of both the theoretical and experimental NDC communities will facilitate this endeavor.

From the experimental perspective, although specific organic molecules with advantageous NDC properties have been identified, they are usually plagued by reliability and stability issues. At the root of the problem may be the nature of molecule-electrode contacts and the experimental conditions, i.e., how the measurements are made, the effects of both of which can easily mask the inherent characteristics of these molecules. Therefore, NDC experiments should incorporate more experimental controls ("knobs") (e.g., temperature, asymmetric contacts, and electrochemical), which will enable researchers to maximize the amount of information they obtain. Meanwhile, because we lack a fundamental physical understanding of NDC, great care must be exercised in attributing mechanisms to, and constructing models of, the observed NDC behaviors, thus dictating the need for critical control experiments. Once we fully understand the NDC mechanism, studies of reliability, including a thorough investigation of all conceivable failure mechanisms, and the subsequent development of optimization steps to correct them, can follow.

On the theoretical side, studies based both on computational methods (such as DFT) and on theoretical modeling have provided a great deal of insight into the possible mechanisms of NDC. However, we feel that despite the progress and the wealth of knowledge that has been amassed, two areas of research in this field are, to some extent, deficient. First, a more detailed comparison between theory and experiments should be provided. While such comparisons have effectively become standard in studies of, e.g., conductance, thermopower, and force spectroscopy, they are still lacking in NDC studies of molecular junctions. This may be due to the need to address fluctuations theoretically, as experimental data is always statistical in nature. However, this task may be difficult, because the fluctuations potentially have multiple origins, such as thermal fluctuations or structural instabilities. Second, as was pointed out above, it is plausible that several NDC mechanisms operate simultaneously. To distinguish between them, theory should attempt to provide more concise predictions and, when possible, suggest "smoking gun" experiments to unambiguously distinguish between the various mechanisms.

None of these tasks is easy, and the experimental and theoretical challenges are considerable. Theory-experiment collaboration in this topic is, in our opinion, essential in overcoming them and elucidating the physical mechanisms governing NDC in molecular junctions. In our view, NDC in molecular junctions is a perfect example where understanding of fundamental physical processes is required to gain any possibility of technological applicability.

**References:**

1. Aviram, A. and M.A. Ratner, *Molecular rectifiers.* Chem. Phys. Lett., 1974. **29**(2): p. 277-283.
2. Chen, J., et al., *Large On-Off Ratios and Negative Differential Resistance in a Molecular Electronic Device.* Science, 1999. **286**(5444): p. 1550-1552.
3. Chen, J., et al., *Room-temperature negative differential resistance in nanoscale molecular junctions.* Appl. Phys. Lett., 2000. **77**(8): p. 1224-1226.
4. Fan, F.-R.F., et al., *Charge Transport through Self-Assembled Monolayers of Compounds of Interest in Molecular Electronics.* J. Am. Chem. Soc., 2002. **124**(19): p. 5550-5560.
5. Kratochvilova, I., et al., *Room temperature negative differential resistance in molecular nanowires.* J. Mater. Chem., 2002. **12**(10): p. 2927-2930.
6. Rawlett, A.M., et al., *Electrical measurements of a dithiolated electronic molecule via conducting atomic force microscopy.* Appl. Phys. Lett., 2002. **81**(16): p. 3043-3045.
7. Xue, Y., et al., *Negative differential resistance in the scanning-tunneling spectroscopy of organic molecules.* Phys. Rev. B, 1999. **59**(12): p. R7852.
8. Ying, H., et al., *Negative differential resistance induced by the Jahn-Teller effect in single molecular coulomb blockade devices.* Computational Materials Science, 2014. **82**: p. 33-36.
9. Mentovich, E.D., et al., *1-Nanometer-Sized Active-Channel Molecular Quantum-Dot Transistor.* Adv. Mater., 2010. **22**(19): p. 2182-2186.
10. Chang, L.L., L. Esaki, and R. Tsu, *Resonant tunneling in semiconductor double barriers.* Appl. Phys. Lett., 1974. **24**(12): p. 593-595.
11. Esaki, L., *New Phenomenon in Narrow Germanium p-n Junctions.* Physical Review, 1958. **109**(2): p. 603-604.
12. Capasso, F., *Resonant Tunneling in Semiconductors: Physics and Applications.* 1991, New York, USA: Plenum Press.
13. Mizuta, H. and T. Tanoue, *The Physics and Applications of Resonant Tunneling Diodes.* 1995, Cambridge, UK: Cambridge University Press.
14. Shahinpoor, M. and H.-J. Schneider, *Intelligent Materials.* 2008, London: Royal Society of Chemistry.
15. Gaudioso, J., L.J. Lauhon, and W. Ho, *Vibrationally Mediated Negative Differential Resistance in a Single Molecule.* Phys. Rev. Lett., 2000. **85**(9): p. 1918-1921.
16. Guisinger, N.P., et al., *Room Temperature Negative Differential Resistance through Individual Organic Molecules on Silicon Surfaces.* Nano Lett., 2003. **4**(1): p. 55-59.
17. Le, J.D., et al., *Negative differential resistance in a bilayer molecular junction.* Appl. Phys. Lett., 2003. **83**(26): p. 5518-5520.
18. Tu, X.W., G. Mikaelian, and W. Ho, *Controlling Single-Molecule Negative Differential Resistance in a Double-Barrier Tunnel Junction.* Phys. Rev. Lett., 2008. **100**(12): p. 126807.
19. Guisinger, N.P., et al., *Observed suppression of room temperature negative differential resistance in organic monolayers on Si(100).* Nanotechnology, 2004. **15**(7): p. S452.
20. Hallbäck, A.-S., B. Poelsema, and H.J.W. Zandvliet, *Negative differential resistance of TEMPO molecules on Si (111).* Appl. Surf. Sci., 2007. **253**(8): p. 4066-4071.
21. Lu, Z.H., et al., *Molecular electronics, negative differential resistance, and resonant tunneling in a poled molecular layer on Al/LiF electrodes having a sharp density of states.* Appl. Phys. Lett., 2004. **85**(2): p. 323-325.
22. Pitters, J.L. and R.A. Wolkow, *Detailed Studies of Molecular Conductance Using Atomic Resolution Scanning Tunneling Microscopy.* Nano Lett., 2006. **6**(3): p. 390-397.
23. Rakshit, T., et al., *Molecules on silicon: Self-consistent first-principles theory and calibration to experiments.* Phys. Rev. B, 2005. **72**(12): p. 125305.
24. Rakshit, T., et al., *Silicon-based Molecular Electronics.* Nano Lett., 2004. **4**(10): p. 1803-1807.
25. Cuevas, J.C. and E. Scheer, *Molecular electronics: an introduction to theory and experiment.* World Scientific series in nanoscience and nanotechnology. 2010, Singapore ; Hackensack, NJ: World Scientific. xix, 703 p.
26. Dubi, Y. and M. Di Ventra, *Colloquium: Heat flow and thermoelectricity in atomic and molecular junctions.* Reviews of Modern Physics, 2011. **83**(1): p. 131.

27. Aradhya, S.V. and L. Venkataraman, *Single-molecule junctions beyond electronic transport.* Nat Nano, 2013. **8**(6): p. 399-410.
28. Migliore, A. and A. Nitzan, *Nonlinear Charge Transport in Redox Molecular Junctions: A Marcus Perspective.* Acs Nano, 2011. **5**(8): p. 6669-6685.
29. Bürkle, M., et al., *Conduction mechanisms in biphenyl dithiol single-molecule junctions.* Phys. Rev. B, 2012. **85**(7): p. 075417.
30. Kang, N., A. Erbe, and E. Scheer, *Observation of negative differential resistance in DNA molecular junctions.* Applied Physics Letters, 2010. **96**(2): p. 023701.
31. Mentovich, E.D., et al., *Multipeak Negative-Differential-Resistance Molecular Device.* Small, 2008. **4**(1): p. 55-58.
32. Reed, M.A., *Molecular-scale electronics.* IEEE Proc., 1999. **87**(4): p. 652-658.
33. Reed, M.A. and J.M. Tour, *Computing with molecules.* Sci. Am., 2000. **282**: p. 86.
34. Xiao, X., et al., *Electrochemical Gate-Controlled Conductance of Single Oligo(phenylene ethynylene)s.* J. Am. Chem. Soc., 2005. **127**(25): p. 9235-9240.
35. Karzazi, Y., J. Cornil, and J.L. Brédas, *Theoretical investigation of the origin of negative differential resistance in substituted phenylene ethynylene oligomers.* Nanotechnology, 2003. **14**(2): p. 165.
36. Grobis, M., et al., *Tuning negative differential resistance in a molecular film.* Appl. Phys. Lett., 2005. **86**(20): p. 204102.
37. Wu, S.W., et al., *Control of Relative Tunneling Rates in Single Molecule Bipolar Electron Transport.* Phys. Rev. Lett., 2004. **93**(23): p. 236802.
38. Choi, B.-Y., et al., *Conformational Molecular Switch of the Azobenzene Molecule: A Scanning Tunneling Microscopy Study.* Phys. Rev. Lett., 2006. **96**(15): p. 156106.
39. Grobis, M., et al., *Tuning negative differential resistance in a molecular film.* Appl. Phys. Lett., 2005. **86**(20): p. -.
40. Chen, L., et al., *Mechanism for Negative Differential Resistance in Molecular Electronic Devices: Local Orbital Symmetry Matching.* Phys. Rev. Lett., 2007. **99**(14): p. 146803.
41. Bevan, K.H., et al., *First-principles nonequilibrium analysis of STM-induced molecular negative-differential resistance on Si(100).* Phys. Rev. B, 2008. **78**(3): p. 035303.
42. Xu, B.Q. and N.J.J. Tao, *Measurement of single-molecule resistance by repeated formation of molecular junctions.* Science, 2003. **301**(5637): p. 1221-1223.
43. Zhou, J., et al., *Measurements of contact specific low-bias negative differential resistance of single metalorganic molecular junctions.* Nanoscale, 2013. **5**(13): p. 5715-5719.
44. Perrin, M.L., et al., *Large negative differential conductance in single-molecule break junctions.* Nat Nano, 2014. **9**(10): p. 830-834.
45. Schwarz, F. and E. Lortscher, *Break-junctions for investigating transport at the molecular scale.* Journal of Physics-Condensed Matter, 2014. **26**(47).
46. Elbing, M., et al., *A single-molecule diode.* Proc. Nat. Acad. Sci. USA, 2005. **102**(25): p. 8815-8820.
47. Lörtscher, E., H.B. Weber, and H. Riel, *Statistical Approach to Investigating Transport through Single Molecules.* Phys. Rev. Lett., 2007. **98**(17): p. 176807.
48. Lörtscher, E., et al., *Reversible and Controllable Switching of a Single-Molecule Junction.* Small, 2006. **2**(8-9): p. 973-977.
49. Galperin, M., M. Ratner, and A. Nitzan, *Hysteresis, switching, and negative differential resistance in molecular junctions: A polaron model.* Nano Letters, 2005. **5**(1): p. 125-130.
50. Galperin, M., M.A. Ratner, and A. Nitzan, *Hysteresis, Switching, and Negative Differential Resistance in Molecular Junctions: A Polaron Model.* Nano Lett., 2004. **5**(1): p. 125-130.
51. Han, J.E., *Nonequilibrium electron transport in strongly correlated molecular junctions.* Phys. Rev. B, 2010. **81**(11): p. 113106.
52. Zhou, J., G. Chen, and B. Xu, *Probing the Molecule-Electrode Interface of Single-Molecule Junctions by Controllable Mechanical Modulations.* J. Phys. Chem. C, 2010. **114**(18): p. 8587-8592.
53. Zhou, J.F., C.L. Guo, and B.Q. Xu, *Electron transport properties of single molecular junctions under mechanical modulations.* Journal of Physics-Condensed Matter, 2012. **24**(16): p. 164029.

54. Zhou, J.F. and B.Q. Xu, *Determining contact potential barrier effects on electronic transport in single molecular junctions.* Applied Physics Letters, 2011. **99**(4): p. 042104.
55. Yeganeh, S., M. Galperin, and M.A. Ratner, *Switching in Molecular Transport Junctions: Polarization Response.* Journal Of The American Chemical Society, 2007. **129**(43): p. 13313-13320.
56. Zazunov, A., D. Feinberg, and T. Martin, *Phonon-mediated negative differential conductance in molecular quantum dots.* Physical Review B, 2006. **73**(11).
57. Di Ventra, M., *Electrical transport in nanoscale systems*. 2008, New York: Cambridge University Press.
58. Meir, Y. and N.S. Wingreen, *Landauer Formula for the Current through an Interacting Electron Region.* Physical Review Letters, 1992. **68**(16): p. 2512-2515.
59. Rutten, B., M. Esposito, and B. Cleuren, *Reaching optimal efficiencies using nanosized photoelectric devices.* Physical Review B, 2009. **80**(23): p. 235122.
60. Muralidharan, B. and S. Datta, *Generic model for current collapse in spin-blockaded transport.* Physical Review B, 2007. **76**(3): p. 035432.
61. Leijnse, M., et al., *Interaction-induced negative differential resistance in asymmetric molecular junctions.* Journal of Chemical Physics, 2011. **134**(10): p. 104107.
62. Einax, M., M. Dierl, and A. Nitzan, *Heterojunction Organic Photovoltaic Cells as Molecular Heat Engines: A Simple Model for the Performance Analysis.* Journal of Physical Chemistry C, 2011. **115**(43): p. 21396-21401.
63. Bonet, E., M.M. Deshmukh, and D.C. Ralph, *Solving rate equations for electron tunneling via discrete quantum states.* Physical Review B, 2002. **65**(4): p. 045317
64. Dubi, Y. and M. Di Ventra, *Thermospin effects in a quantum dot connected to ferromagnetic leads* Physical Review B, 2009. **80**(11): p. 081302.
65. Braun, M., J. Konig, and J. Martinek, *Theory of transport through quantum-dot spin valves in the weak-coupling regime.* Physical Review B, 2004. **70**(19).
66. Koller, S., et al., *Density-operator approaches to transport through interacting quantum dots: Simplifications in fourth-order perturbation theory.* Physical Review B, 2010. **82**(23).
67. Esposito, M. and M. Galperin, *Transport in molecular states language: Generalized quantum master equation approach.* Physical Review B, 2009. **79**(20).
68. Esposito, M. and M. Galperin, *Self-Consistent Quantum Master Equation Approach to Molecular Transport.* Journal of Physical Chemistry C, 2010. **114**(48): p. 20362-20369.
69. Harbola, U., M. Esposito, and S. Mukamel, *Quantum master equation for electron transport through quantum dots and single molecules.* Physical Review B, 2006. **74**(23).
70. Elstner, M., et al., *Self-consistent-charge density-functional tight-binding method for simulations of complex materials properties.* Physical Review B, 1998. **58**(11): p. 7260-7268.
71. Di Carlo, A., et al., *Theoretical tools for transport in molecular nanostructures.* Physica B-Condensed Matter, 2002. **314**(1-4): p. 86-90.
72. Aradi, B., B. Hourahine, and T. Frauenheim, *DFTB+, a sparse matrix-based implementation of the DFTB method.* Journal of Physical Chemistry A, 2007. **111**(26): p. 5678-5684.
73. Schmitteckert, P., *The dark side of DFT based transport calculations.* Physical Chemistry Chemical Physics, 2013. **15**(38): p. 15845-15849.
74. Dubi, Y., *Possible origin of thermoelectric response fluctuations in single-molecule junctions.* New Journal of Physics, 2013. **15**: p. 105004.
75. Vignale, G. and M. Di Ventra, *Incompleteness of the Landauer formula for electronic transport.* Physical Review B, 2009. **79**(1): p. 014201.
76. Mujica, V. and M.A. Ratner, *Current-voltage characteristics of tunneling molecular junctions for off-resonance injection.* Chemical Physics, 2001. **264**(3): p. 365-370.
77. Mujica, V., A.E. Roitberg, and M. Ratner, *Molecular wire conductance: Electrostatic potential spatial profile.* Journal of Chemical Physics, 2000. **112**(15): p. 6834-6839.
78. Cheraghchi, H. and K. Esfarjani, *Negative differential resistance in molecular junctions: Application to graphene ribbon junctions.* Physical Review B, 2008. **78**(8): p. 085123.

79. Dalgleish, H. and G. Kirczenow, *Interface states, negative differential resistance, and rectification in molecular junctions with transition-metal contacts.* Physical Review B, 2006. **73**(24): p. 245431.
80. Dalgleish, H. and G. Kirczenow, *A new approach to the realization and control of negative differential resistance in single-molecule nanoelectronic devices: Designer transition metal-thiol interface states.* Nano Letters, 2006. **6**(6): p. 1274-1278.
81. Emberly, E.G. and G. Kirczenow, *Current-driven conformational changes, charging, and negative differential resistance in molecular wires.* Phys. Rev. B, 2001. **64**(12): p. 125318.
82. Garcia-Suarez, V.M. and J. Ferrer, *Nonequilibrium transport response from equilibrium transport theory.* Physical Review B, 2012. **86**(12): p. 125446
83. Geng, H., et al., *Molecular design of negative differential resistance device through intermolecular interaction.* Journal of Physical Chemistry C, 2007. **111**(51): p. 19098-19102.
84. Huang, J., et al., *Iron-phthalocyanine molecular junction with high spin filter efficiency and negative differential resistance.* Journal of Chemical Physics, 2012. **136**(6): p. 064707.
85. Karzazi, Y., J. Cornil, and J.L. Brédas, *Negative Differential Resistance Behavior in Conjugated Molecular Wires Incorporating Spacers: A Quantum-Chemical Description.* J. Am. Chem. Soc., 2001. **123**(41): p. 10076-10084.
86. Qiu, M., et al., *Transport Properties of a Squeezed Carbon Monatomic Ring: A Route to a Negative Differential Resistance Device.* Journal of Physical Chemistry C, 2011. **115**(23): p. 11734-11737.
87. Thygesen, K.S., *Impact of exchange-correlation effects on the IV characteristics of a molecular junction.* Physical Review Letters, 2008. **100**(16): p. 166804.
88. Wang, Y. and H.-P. Cheng, *Electronic and transport properties of azobenzene monolayer junctions as molecular switches.* Physical Review B, 2012. **86**(3): p. 035444.
89. Xia, C.J., et al., *Large negative differential resistance in a molecular device with asymmetric contact geometries: A first-principles study.* Physica E-Low-Dimensional Systems & Nanostructures, 2011. **43**(8): p. 1518-1521.
90. Zimbovskaya, N.A. and M.R. Pederson, *Negative differential resistance in molecular junctions: Effect of the electronic structure of the electrodes.* Physical Review B, 2008. **78**(15): p. 153105.
91. Verzijl, C.J.O., J.S. Seldenthuis, and J.M. Thijssen, *Applicability of the wide-band limit in DFT-based molecular transport calculations.* Journal of Chemical Physics, 2013. **138**(9): p. 094102.
92. Soler, J.M., et al., *The SIESTA method for ab initio order-N materials simulation.* Journal of Physics-Condensed Matter, 2002. **14**(11): p. 2745-2779.
93. Stokbro, K., et al., *Semiempirical model for nanoscale device simulations.* Physical Review B, 2010. **82**(7): p. 075420.
94. Pati, R., M. McClain, and A. Bandyopadhyay, *Origin of negative differential resistance in a strongly coupled single molecule-metal junction device.* Physical Review Letters, 2008. **100**(24): p. 246801
95. Galperin, M., A. Nitzan, and M.A. Ratner, *The non-linear response of molecular junctions: the polaron model revisited.* Journal of Physics-Condensed Matter, 2008. **20**(37): p. 374107.
96. Galperin, M., M.A. Ratner, and A. Nitzan, *Molecular transport junctions: vibrational effects.* Journal of Physics-Condensed Matter, 2007. **19**(10): p. 103201.
97. Entin-Wohlman, O., Y. Imry, and A. Aharony, *Three-terminal thermoelectric transport through a molecular junction.* Physical Review B, 2010. **82**(11): p. 115314.
98. Alexandrov, A.S. and A.M. Bratkovsky, *Memory effect in a molecular quantum dot with strong electron-vibron interaction.* Physical Review B, 2003. **67**(23).
99. Mozyrsky, D., M.B. Hastings, and I. Martin, *Intermittent polaron dynamics: Born-Oppenheimer approximation out of equilibrium.* Physical Review B, 2006. **73**(3).
100. Pistolesi, F., Y.M. Blanter, and I. Martin, *Self-consistent theory of molecular switching.* Physical Review B, 2008. **78**(8).
101. D'Amico, P., et al., *Charge-memory effect in a polaron model: equation-of-motion method for Green functions.* New Journal of Physics, 2008. **10**.
102. Koch, J., F. von Oppen, and A.V. Andreev, *Theory of the Franck-Condon blockade regime.* Physical Review B, 2006. **74**(20).

103. Donarini, A., A. Yar, and M. Grifoni, *Vibration induced memory effects and switching in ac-driven molecular nanojunctions.* European Physical Journal B, 2012. **85**(9).
104. Koch, J., et al., *Thermopower of single-molecule devices.* Physical Review B, 2004. **70**(19).
105. Koch, J. and F. von Oppen, *Franck-Condon blockade and giant Fano factors in transport through single molecules.* Physical Review Letters, 2005. **94**(20).
106. Koch, J., et al., *Current-induced nonequilibrium vibrations in single-molecule devices.* Physical Review B, 2006. **73**(15).
107. Ryndyk, D.A., et al., *Charge-memory polaron effect in molecular junctions.* Physical Review B, 2008. **78**(8).
108. La Magna, A. and I. Deretzis, *Phonon driven nonlinear electrical behavior in molecular devices.* Physical Review Letters, 2007. **99**(13): p. 136404.
109. Koch, T., et al., *Nonequilibrium transport through molecular junctions in the quantum regime.* Physical Review B, 2011. **84**(12): p. 125131.
110. Han, J., *Nonequilibrium electron transport in strongly correlated molecular junctions.* Physical Review B, 2010. **81**(11): p. 113106.
111. Hartle, R. and M. Thoss, *Resonant electron transport in single-molecule junctions: Vibrational excitation, rectification, negative differential resistance, and local cooling.* Physical Review B, 2011. **83**(11): p. 115414
112. Kuznetsov, A.M., *Negative differential resistance and switching behavior of redox-mediated tunnel contact.* Journal of Chemical Physics, 2007. **127**(8): p. 084710
113. Perfetto, E. and G. Stefanucci, *Image charge effects in the nonequilibrium Anderson-Holstein model.* Physical Review B, 2013. **88**(24): p. 245437
114. Park, J., et al., *Coulomb blockade and the Kondo effect in single-atom transistors.* Nature, 2002. **417**(6890): p. 722-725.
115. Kubatkin, S., et al., *Single-electron transistor of a single organic molecule with access to several redox states.* Nature, 2003. **425**(6959): p. 698-701.
116. Refaely-Abramson, S., R. Baer, and L. Kronik, *Fundamental and excitation gaps in molecules of relevance for organic photovoltaics from an optimally tuned range-separated hybrid functional.* Physical Review B, 2011. **84**(7): p. 075144.
117. Migliore, A. and A. Nitzan, *Irreversibility and Hysteresis in Redox Molecular Conduction Junctions.* Journal of the American Chemical Society, 2013. **135**(25): p. 9420-9432.
118. Migliore, A., P. Schiff, and A. Nitzan, *On the relationship between molecular state and single electron pictures in simple electrochemical junctions.* Physical Chemistry Chemical Physics, 2012. **14**(40): p. 13746-13753.
119. Nitzan, A., A. Migliore, and P. Schiff, *Redox molecular junctions: Properties and functionalities.* Abstracts of Papers of the American Chemical Society, 2012. **244**.
120. White, A.J., et al., *Quantum transport with two interacting conduction channels.* Journal of Chemical Physics, 2013. **138**(17): p. 174111.
121. Bekenstein, Y., et al., *Periodic negative differential conductance in a single metallic nanocage.* Physical Review B, 2012. **86**(8): p. 085431.
122. Kaasbjerg, K. and K. Flensberg, *Image charge effects in single-molecule junctions: Breaking of symmetries and negative-differential resistance in a benzene single-electron transistor.* Physical Review B, 2011. **84**(11): p. 115457.
123. Parida, P., S. Lakshmi, and S.K. Pati, *Negative differential resistance in nanoscale transport in the Coulomb blockade regime.* Journal of Physics-Condensed Matter, 2009. **21**(9): p. 095301.
124. Parida, P., S. Pati, and A. Painelli, *Negative differential conductance in nanojunctions: A current constrained approach.* Physical Review B, 2011. **83**(16): p. 165404.
125. Tsai, M.H. and T.H. Lu, *Electronic and transport properties of a molecular junction with asymmetric contacts.* Nanotechnology, 2010. **21**(6): p. 065203.
126. Thygesen, K.S. and A. Rubio, *Renormalization of Molecular Quasiparticle Levels at Metal-Molecule Interfaces: Trends across Binding Regimes.* Physical Review Letters, 2009. **102**(4): p. 046802.

127. Dzhioev, A.A. and D.S. Kosov, *Solvent-induced current-voltage hysteresis and negative differential resistance in molecular junctions.* Physical Review B, 2012. **85**(3).
128. Dubi, Y., *Dynamical coupling and negative differential resistance from interactions across the molecule-electrode interface in molecular junctions.* Journal of Chemical Physics, 2013. **139**(15): p. 154710.